# Quantum Interference, Graphs, Walks, and Polynomials

Yuta Tsuji[1)], Ernesto Estrada[2)], Ramis Movassagh[3)], Roald Hoffmann*[4)]

*1) Institute for Materials Chemistry and Engineering and IRCCS, Kyushu University, Nishi-ku, Fukuoka 819-0395, Japan*

*2) Department of Mathematics and Statistics, University of Strathclyde, 26 Richmond Street, Glasgow G11HX, United Kingdom*

*3) IBM Research, MIT-IBM A.I. Lab, Cambridge MA 02142, USA*

*4) Department of Chemistry and Chemical Biology, Cornell University, Ithaca, NY 14853-1301, USA*

***ABSTRACT:*** *In this paper we explore quantum interference in molecular conductance from the point of view of graph theory and walks on lattices. By virtue of the Cayley-Hamilton theorem for characteristic polynomials and the Coulson-Rushbrooke pairing theorem for alternant hydrocarbons, it is possible to derive a finite series expansion of the Green's function for electron transmission in terms of the odd powers of the vertex adjacency matrix or Hückel matrix. This means that only odd-length walks on a molecular graph contribute to the conductivity through a molecule. Thus, if there are only even-length walks between two atoms, quantum interference is expected to occur in the electron transport between them. However, even if there are only odd-length walks between two atoms, a situation may come about where the contributions to the QI of some odd-length walks are cancelled by others, leading to another class of quantum interference. For non-alternant hydrocarbons, the finite Green's function expansion may include both even and odd powers. Nevertheless, QI can in some circumstances come about for non-alternants, from cancellation of odd and even-length walk terms. We report some progress, but not a complete resolution of the problem of understanding the coefficients in the expansion of the Green's function in a power series of the adjacency matrix, these coefficients being behind the cancellations that we have mentioned. And we introduce a perturbation theory for transmission as well as some potentially useful infinite power series expansions of the Green's function.*

**CONTENTS**

## 1. INTRODUCTION

Since any molecule is a quantum object, the quantum superposition principle leads to non-negligible quantum effects on electron transport through a single molecule.[1,2] A striking, non-classical exemplification of this is in destructive quantum interference (QI), where a dramatic diminution in molecular conductance is observed – for certain pathways, certain connections, not others.[3,4,5] A challenge to chemists, who have learned to understand other quantum effects, is to develop an intuition that partakes of both classical chemical ideas and quantum mechanics in understanding and manipulating QI features in molecular electronics.[6]

The main goal of this review is to explore how one can relate a path or walk on a graph, representing a molecule, to the series expansion of the Green's function for the system. And through that route we will establish a strong link between the connectivity of a molecule and QI. There are different ways to account for this connection between walks and QI. We could also approach the field through the Dyson equation and Feynman paths, sophisticated methods to describe resonant conduction.[7] This methodology is useful when the Fermi energy of the electrode matches at least one of the MO levels of a molecule bridging the electrodes. Although QI may occur in the resonant conduction regime,[8] it often occurs in off-resonant conduction,[9] where the Fermi energy of the electrode is located in-between the HOMO and LUMO of the bridged molecule.

Given the mathematical complexity of some of the materials, we adopt a pedagogical style in most of the review, so as to make understandable the fundamentals of this important connection between walks on graphs and QI. The work can be read in several ways and we first start by giving the reader a guide on how to navigate this work.

## 2. HOW TO READ THIS REVIEW?

We start this review by describing the main contributions reported in the literature to explain QI in molecules in §3. Then, we continue by explaining the relation between the Green's function and graph-theoretic path counting, which can be developed to describe off-resonant conduction and is a field explored earlier by Estrada.

We now summarize the contents of each subsequent section so that the reader can decide which sections to read in detail.

§4 is an introductory section in which the basics of chemical graph theory for π conjugated molecules are detailed. The adjacency matrix is introduced and is related to the Hückel Hamiltonian matrix. Also, the bipartite graph, which corresponds to alternant hydrocarbons in chemistry, is explained. For more detailed graph theoretic terminology, the reader may refer to a useful review by Essam and Fisher, which makes many connections between graph theory and physics.[10] Expert readers, who may be familiar with basic graph theory and the Hückel method, can skip this section.

§5 is one of the most important sections in this review. We introduce the Green's

function and then connect it with the adjacency matrix. We use an important symmetric feature found in the eigenspectrum of the bipartite graphs or alternant hydrocarbons, namely the Coulson-Rushbrooke pairing theorem,[11] in conjunction with the Cayley-Hamilton theorem, to derive a finite power series expansion of the Green's function in terms of the adjacency matrix. Based on the derived series expansion, we show how one can connect electron conduction or QI with walks on a graph.

The power series expansion of the Green's function derived in §5 is only applicable to closed-shell alternant hydrocarbons, or non-singular bipartite graphs in graph theoretic terminology. In §6, we describe how the Green's function expansion looks when applied to non-alternant hydrocarbons, or non-bipartite graphs. In the subsections 6.1, 6.2, and 6.3, we use three non-alternant examples, namely fulvene, [3]radialene, and azulene. Readers who have interest in electron transport through nonalternants should find this section useful.

In the adjacency-matrix-based power series expansions of the Green's function for alternant and non-alternant hydrocarbons (bipartite and non-bipartite graphs) derived in §5 and §6, the coefficients in the power series are calculated from the orbital energies of the molecule or the eigenvalues of the graph. In §7, we delve into the origin of the coefficients, reviewing three different approaches, namely Sachs graphs, Newton's identities, and Hosoya's non-adjacent or edge independence numbers. Though the contents of this section are quite mathematical, as one proceeds through the section one returns along the way to chemistry, as one can figure out how to relate the coefficients with the moments of molecular-energy spectra as well as the number of radical valence-bond structures. This section details work in progress; we make the reader aware of some interesting challenges that remain to be solved.

Though the Green's function method is a standard method to calculate molecular conduction properties, scattering theory and source-and-sink potential theory are also very useful. In §8, we review these two methods and show important determinantal equations, through which we arrive at the characteristic polynomial and clarify the relation to our Green's function approach. Also, we provide an insight into the coefficients of the characteristic polynomial that can enhance chemists' understanding of Sachs graphs and Hosoya's non-adjacency number concept.

In §9, we continue to work on the important coefficients, but here we develop our own methodology, which we call pairwise bond orders or pairwise bond interactions. An interesting question about formulating through-bond and through-space interactions based on graph-theoretic Hückel methods emerges.

Perturbation theory has proven of immense utility in providing chemistry with a language for formulating interpretations. In §10 and §11, we turn to a perturbation theory for transmission, which is intimately related to the other parts of this review, e.g. graph theory, walks, characteristic polynomials and power series expansions. To begin with, in §10, we review perturbation theory in chemistry, and then we inspect how it works in a graph theoretical way of thinking, where we can regard non-alternant hydrocarbons, or non-bipartite graphs, as a perturbed system generated from an unperturbed alternant system, or bipartite graph. We see how QI features in non-alternants are affected by perturbation.

Earlier in the review, in §5, we will have derived a finite power series expansion of the Green's function. In section §11, we examine a couple of infinite power series expansions utilizing a Neumann series or the binominal theorem. We first discuss the problem of convergence. A perturbation matrix is introduced. For one thing, it provides a bridge between alternant and non-alternant hydrocarbons; for another, it ensures convergence in the expansion. This section is very mathematical, but we are sure that such elaborations will be useful for understanding electron transfer in two important subclasses of π conjugated molecules. Finally, in §12, we summarize and conclude our review. Enjoy the journey!

## 3. STATE OF THE ART IN THE THEORY OF QI

The last 10 years have seen the emergence of a wide variety of interpretations of QI. For example, Solomon and co-workers developed a method to examine the phase of electron transmission[12] as well as a clear visualization of QI using local atom-to-atom transmission.[13] Nozaki and co-workers introduced a parabolic diagram approach to clearly visualize the conditions for the occurrence of QI.[14] Markussen, Stadler, and Thygesen proposed a simple and useful diagrammatic method that provides a direct link between QI and the topology of various π-conjugated systems;[15] recently, this method has been further developed by Pedersen et al.[16] These diagrammatic approaches to QI connect to seminal work by Stadler, Ami, Joachim, and Forshaw.[17] They used an electron scattering formalism based on a topological Hückel description, deriving a graphical method which allows a quick assessment whether QI occurs or not. Their methodology will be further reviewed in section 8.

Molecular orbitals (MOs), which have become part of the toolkit of all chemists, clearly play a role in understanding and applying QI. For example, Yoshizawa and Tada pointed out that the amplitude and phase of frontier MOs play a crucial role in the manifestation of QI.[18] Stadler and co-workers have indicated the limited applicability of the orbital rule to predict QI in alternant hydrocarbons on the basis of Larsson's formula.[19] Ernzerhof used "device orbital theory" to predict QI.[20] Markussen and co-workers obtained a simple orbital-based explanation of QI by transforming the frontier MOs into localized MOs.[21] Bürkle and co-workers developed a two level model and found that interorbital coupling plays the decisive role for QI effects.[22]

Still other concepts from theoretical chemistry have been fruitfully correlated with QI. For example, Stuyver and co-workers have pointed out the relation of QI to the atom-atom polarizability,[23] Pauling's bond order,[24] the number of Kekulé structures,[24] and "electron pushing", the curly arrow formalism widely utilized in organic chemistry.[25] Hosoya independently noticed the implication of the curly arrow scheme in QI.[26] In association with Strange and Solomon, we explored the close relation between QI and diradical existence.[27] Nakano and co-workers pointed out the relationship of QI with the remarkable phenomenon of singlet fission.[28]

It is intuitive to think of a pathway along which electrons flow in a molecule, an aggregation of molecules, or solids in the process of electron transfer or transport. Beratan and co-workers established a tunneling pathway model to analyze nonadiabatic electron transfer inside biomolecules, such as proteins and DNAs, decomposing the total transmission into the contributions from paths such as covalent bonds, hydrogen bonds, and van der Waals contacts.[29,30,31,32] In this model, one can identify the most facile electronic-coupling routes between the electron-donor and acceptor. The method neglects the effects of interferences between multiple tunneling pathways, because the model only employs semi-empirical parameters and is not based on wave functions, which bear the information about quantum mechanical superposition. Newton and co-workers developed another pathway analysis scheme, in which effective transfer integrals can be decomposed into additive contributions from individual pathways, both through-space and through-bond, using perturbation theory based on a localized orbital basis represented by natural bond orbitals.[33,34] Marcus and co-workers introduced a combined artificial intelligence-superexchange methodology, which employs the overlap integral between atomic orbitals.[35,36] In these models, some interference effects can be included.

One might be tempted to attribute the QI feature of molecular conductance to the outcome of interfering electron waves passing through two different pathways in real space[37] in analogy with a Mach–Zehnder interferometer[38] or a double slit experiment (cf. free-electron network model[39]). For example, in the case of electron transport through a *meta*-substituted benzene ring, which is a canonical example of QI, one can clearly recognize two pathways: a shorter one and a longer one. And one intuitive interpretation of the resulting QI is that the phase shift of the transporting electron waves due to the different path lengths leads to destructive interference between them.[40]

Recently, there has been an active discussion as to whether QI features are caused by electron-wave interference between different paths through space or that between different MOs in energy space.[41,42] Lambert and co-workers derived analytical formulae describing electron transport through single-path and multi-path structures, demonstrating that QI does not require the presence of physically different paths, as interference might be caused by scattering from nodal impurity sites and connections to external leads.[43] Nozaki and Toher[44] also addressed this problem by investigating the evolution of the transmission dip upon the attenuation of the resonance integral of one bond, which corresponds to blocking of one electron-transport pathway. Nevertheless, the transmission antiresonance remains intact. Thus, they argued that the analogy between the classical double-slit experiment and QI is not correct.

Further debate ensued.[45,46]

If one wants to understand QI effects, it may also make sense to use the formalism of Feynman paths.[47] QI features might be interpreted as the consequence of the interference between electron wave-functions through different Feynman paths.[48] By means of the Feynman path idea, one may obtain qualitative physical insight into why the Green's function represents the propagation of electrons.[49] To put it another way, a physical picture of molecular conductance can be given by a sum over a huge number of Feynman paths.[50]

One can easily find in the literature examples of how the Feynman path formalism can enhance understanding of a wide range of conduction problems. For example, Lee investigated conductance fluctuations in disordered metals, where the important (and numerous) Feynman paths are found to be random walks which cover much of the sample.[50] Datta and co-workers demonstrated that the Feynman path formalism has an intimate relation to a scattering-matrix approach developed to calculate the conductance of disordered systems.[51] Gong et al. carried out a Feynman path analysis of electron transport though a parallel double quantum dot (QD) structure, finding that there are infinite electron transmission paths which contribute to a Fano interference.[48,52] Huo pointed out that one can use the Feynman path analysis to acquire a qualitative understanding of the physical nature of QIs, such as the Kondo resonance and Fano antiresonance, which can be observed in the conductance spectrum of a laterally coupled carbon-nanotube QD system.[53] Since the energy levels of QDs are discrete, electron transport through QDs is thought to occur by resonant tunneling, where the energy of an incident electron coincides with an eigenenergy of the QD. The studies by Gong et al. and Huo clearly show that the Feynman path analysis is an effective tool for understanding the propagation of electrons in the resonant tunneling regime. Problems may arise in the use of Feynman paths in actual computations of conductance because of the infinitely multiple paths involved, though higher-order Feynman paths are not likely to play a significant role.

Feynman paths are less familiar to chemists than physicists, yet their utility in quantum mechanical problems of some generality makes it clear that it is worthwhile for chemists to learn the formalism attached. We will devote some space to introducing the subject in the Supporting Information (SI) to this paper.

The relationship between graph theory and quantum chemistry is more direct, and long-standing.[54,55] The connection is well-established as far as electronic structure goes. Perhaps there was a time earlier, a time when it was important to establish a tie between valence bond (VB) and MO theory, when the connection between graph theory and electronic structure was more central to the field than it is today. But the relationship of density of states (DOS) to moments of DOS, and through them to walks on a lattice, has played an important part in relating geometry, therefore directly molecular and solid state structure, to the relative energy of various structures.[56,57] Since the DOS can also be obtained through a Green's function technique, one can correlate the Green's function with the moments through a power series expansion.[58,59]

Since there is a good correspondence between the Hückel or tight-binding method and graph theory (which we will review), QI can be understood in the context of walks on graphs and connectivity of atoms in a molecule. For example, Estrada has developed a graph-theoretic path-counting method to predict QI based on the Cayley-Hamilton and binomial theorems.[60] Fowler, Pickup, and their co-workers developed a formulation of the electron-transmission function in terms of characteristic polynomials, deriving another simple selection rule to predict QI by counting the number of non-bonding levels in the molecular graph and some vertex-deleted subgraphs.[61,62] They further found links between their selection rule and various chemical concepts, such as Kekulé structures and bond orders.[63] Their polynomial-based approach to transmission will be further reviewed in section 8. Recently, a Green's function approach to quantum graphs has been published by Andrade and co-workers.[64]

## 4. GRAPH THEORY FOR π SYSTEMS

In this review we will consider electron transport phenomena through π-conjugated hydrocarbons. This is a subclass of all molecules, for sure, but an important one. In alternant hydrocarbons, a further subdivision of these, all the carbon atoms belong either to a starred set or an unstarred set so that no atom of one set is adjacent to another atom of the same set (see Scheme 1a).

***Scheme 1.*** *(a) Examples of alternant and non-alternant hydrocarbons with star marks. The dashed circle indicates the region of frustration of the starring scheme for a non-alternant. (b) To illustrate an example of the adjacency matrix, that for fulvene is shown.*

(a)

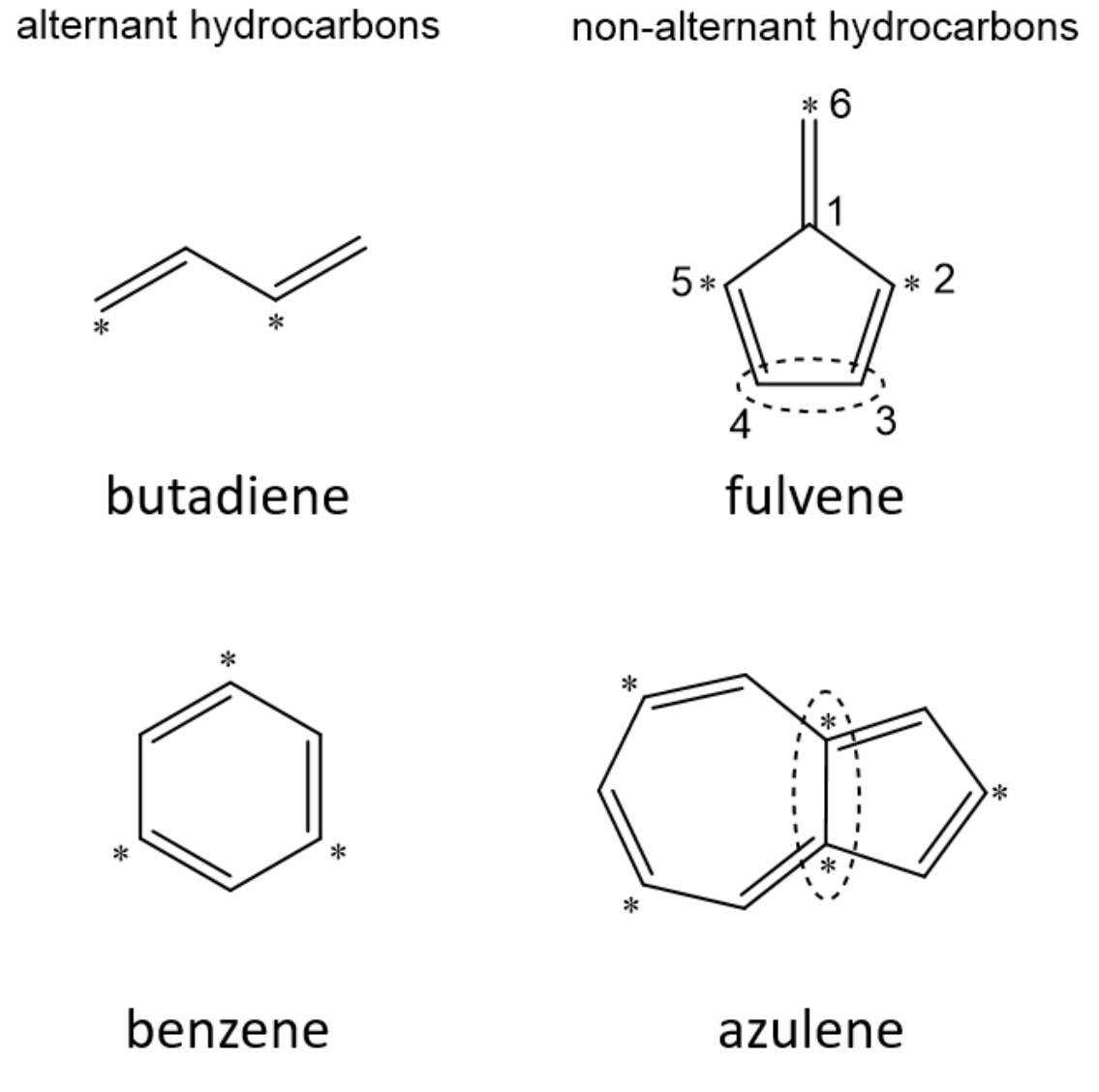

(b)

$$\mathbf{A} = \begin{array}{c} \\ 1 \\ 2 \\ 3 \\ 4 \\ 5 \\ 6 \end{array} \begin{array}{c} \begin{array}{cccccc} 1 & 2 & 3 & 4 & 5 & 6 \end{array} \\ \begin{bmatrix} 0 & 1 & 0 & 0 & 1 & 1 \\ 1 & 0 & 1 & 0 & 0 & 0 \\ 0 & 1 & 0 & 1 & 0 & 0 \\ 0 & 0 & 1 & 0 & 1 & 0 \\ 1 & 0 & 0 & 1 & 0 & 0 \\ 1 & 0 & 0 & 0 & 0 & 0 \end{bmatrix} \end{array}$$

fulvene

Let $\mathbb{G} = (V, E)$ be a molecular graph with a vertex adjacency matrix **A**. We represent the hydrogen-deleted skeleton of a π-conjugated hydrocarbon in the form of a graph, such that every $sp^2$ carbon atom is represented by a node $i \in V$ and every σ-bond between two $sp^2$ carbon atoms $i$ and $j$ is represented by an edge $(i,j) \in E$. $\mathbb{G}$ for alternant hydrocarbons is called a bipartite graph. The vertex adjacency matrix **A** is the matrix with entries +1 for pairs of atoms (indices) connected=adjacent, and 0 for those not connected=bonded. In Scheme 1b we show, by way of illustration, **A** for a nonalternant, fulvene.

We adopt the tight-binding/Hückel Hamiltonian matrix **H** for the description of π-systems.[65,66] Assuming that all the carbon atoms have the same on-site Coulomb energy, whose value can be set to zero without loss of generality, all the diagonal elements of **H** are equal to zero and we can write the matrix as

$$\mathbf{H} = \beta\mathbf{A}, \quad (1)$$

where $\beta$ is the resonance/hopping integral between adjacent carbon 2pπ orbitals. If $\beta$ is used as the unit of energy, the Hamiltonian is further simplified as

$$\mathbf{H} = \mathbf{A}. \quad (2)$$

So these two matrices/operators correlate directly at the simple Hückel level.[66] We should note that because $\beta$ is negative, the energy levels $\varepsilon_j$ of a conjugated molecule correspond to the negative of the eigenvalues, i.e., $\varepsilon_j = -\lambda_j$.[67] MO coefficients and MO energies translate into eigenvectors and eigenvalues of the adjacency matrix, which are dependent on the connectivity of atoms (vertices) in a molecule (graph), namely topology.

## 5. UNDERSTANDING OF THE GREEN'S FUNCTION BASED ON WALKS ON A GRAPH

In this section, we will discuss how the connectivity of atoms can be related to the conductivity of a single molecule. Indeed, graph theory has been utilized to describe the conductivity on various scales, from macroscopic electrical circuits[68] to mesoscopic amorphous materials.[69] The last decade has witnessed a surge in contributions from graph theory to theories of conductivity on a single-molecular scale.[60, 70, 71, 72, 73, 74] For electron-transport calculations, one may use what is called source-and-sink potential (SSP) method,[75,76] which has significantly aided the development of the graph-theoretic approaches to molecular conductance. And we have also benefitted from it.[77] However, in this paper, we follow an alternative route, the Green's function method, perhaps a more common protocol in the molecular electronics community.

The Green's function for the molecular graph $\mathbb{G}$ represented by **A** can be written as

$$\mathbf{G}(E) = [E\mathbf{I} - \mathbf{A}]^{-1}. \quad (3)$$

Note that the symbol **G** for Green's function is not to be confused with $\mathbb{G}$ for graph. If we assume the Fermi energy is located at $E = 0$, the Green's function at the Fermi level takes on a very simple form,

$$\mathbf{G}(E_F) = -\mathbf{A}^{-1}. \tag{4}$$

This equation holds true for electron transport in the off-resonant regime, where the Fermi energy ($E = 0$) does not match any of the eigenvalues of **A**. To put it another way, **A** is not allowed to have any zero eigenvalues, if **A** is to be invertible. In chemical terminology, this condition can be expressed alternatively as follows: The molecule represented by a molecular graph $\mathbb{G}$ should not be a radical (e.g., diradical, tetraradical, or higher-order radical) as radicals have one or more zero eigenvalues of the adjacency matrix. In mathematical terminology, the adjacency matrix **A** of the graph $\mathbb{G}$ has to be full rank. In graph theoretic terminology, the graph $\mathbb{G}$ has to be a non-singular graph.

Let us think about the eigenvalues of the Hamiltonian/adjacency matrix. The characteristic polynomial for **A** is resolved into factors[78] by using its eigenvalues $\varepsilon_i$,

$$p(\lambda) = \det(\mathbf{A} - \lambda\mathbf{I}) = (\lambda - \varepsilon_1)(\lambda - \varepsilon_2)(\lambda - \varepsilon_3)\cdots(\lambda - \varepsilon_N). \tag{5}$$

This is equivalent to the direct expansion of a Hückel secular determinant, which has been found to have a connection with a variety of physicochemical quantities of molecules, such as charge densities, bond orders, total energies, and polarizabilities.[79,80] Since the systems we are considering are limited to even alternant hydrocarbons whose molecular graph is a non-singular bipartite graph, we can use what in theoretical chemistry is called the pairing theorem of Coulson and Rushbrooke.[11] This theorem states that if the molecule has an energy level of $\varepsilon_i$, then $-\varepsilon_i$ is also an eigenvalue of the Hamiltonian/adjacency matrix. If the molecule/graph consists of $N$ atoms/vertices, there are $N/2$ positive eigenvalues and $N/2$ negative eigenvalues. And they are paired. So the characteristic polynomial can be simplified as

$$p(\lambda) = (\lambda - \varepsilon_1)(\lambda + \varepsilon_1)(\lambda - \varepsilon_2)(\lambda + \varepsilon_2)\cdots(\lambda - \varepsilon_{N/2})(\lambda + \varepsilon_{N/2}) = \prod_{i=1}^{N/2}(\lambda^2 - \varepsilon_i^2) \tag{6}$$

To obtain an expression for the Green's function or the inverse matrix by means of the characteristic polynomial, one can use the Cayley-Hamilton theorem.[81] This remarkable and useful theorem states that an arbitrary $N \times N$ matrix **A** satisfies its own characteristic equation, namely $p(\mathbf{A}) = \mathbf{0}$, where $\mathbf{0}$ is the zero matrix. When this theorem is applied to eq. 6, it leads to

$$p(\mathbf{A}) = \prod_{i=1}^{N/2}(\mathbf{A}^2 - \varepsilon_i^2\mathbf{I}) = \mathbf{0}. \tag{7}$$

To get some additional insight into eq. 7, we write down the explicit expressions for the case of $N = 2$, 4, and 6 in the following. For $N = 2$,

$$p(\mathbf{A}) = \mathbf{A}^2 - \varepsilon_1^2\mathbf{I} = \mathbf{0}. \tag{8}$$

For $N = 4$,

$$p(\mathbf{A}) = \mathbf{A}^4 - (\varepsilon_1^2 + \varepsilon_2^2)\mathbf{A}^2 + \varepsilon_1^2\varepsilon_2^2\mathbf{I} = \mathbf{0}. \tag{9}$$

For $N = 6$,

$$p(\mathbf{A}) = \mathbf{A}^6 - (\varepsilon_1^2 + \varepsilon_2^2 + \varepsilon_3^2)\mathbf{A}^4 + (\varepsilon_1^2\varepsilon_2^2 + \varepsilon_1^2\varepsilon_3^2 + \varepsilon_2^2\varepsilon_3^2)\mathbf{A}^2 - \varepsilon_1^2\varepsilon_2^2\varepsilon_3^2\mathbf{I} = \mathbf{0}. \tag{10}$$

One can derive a generalized expression for eq. 7 in a matrix polynomial form, as follows:

$$p(\mathbf{A}) = \mathbf{A}^N - \left(\sum_i^{N/2} \varepsilon_i^2\right)\mathbf{A}^{N-2} + \left(\sum_{i \neq j}^{N/2} \varepsilon_i^2\varepsilon_j^2\right)\mathbf{A}^{N-4} + \cdots + (-1)^{N/2}\left(\prod_i^{N/2} \varepsilon_i^2\right)\mathbf{I} = \mathbf{0}. \tag{11}$$

By multiplying the expression by the inverse matrix of **A**, we can connect this equation to the expression of the Green's function in eq. 4 as follows:

$$\mathbf{A}^{N-1} - \left(\sum_i^{N/2} \varepsilon_i^2\right)\mathbf{A}^{N-3} + \left(\sum_{i \neq j}^{N/2} \varepsilon_i^2\varepsilon_j^2\right)\mathbf{A}^{N-5} + \cdots + (-1)^{N/2}\left(\prod_i^{N/2} \varepsilon_i^2\right)\mathbf{A}^{-1} = \mathbf{0} \tag{12}$$

Since the Green's function **G** is equal to $-\mathbf{A}^{-1}$, it can be written as

$$\mathbf{G} = \frac{(-1)^{N/2}}{\prod_i^{N/2} \varepsilon_i^2}\left[\mathbf{A}^{N-1} - \left(\sum_i^{N/2} \varepsilon_i^2\right)\mathbf{A}^{N-3} + \left(\sum_{i \neq j}^{N/2} \varepsilon_i^2\varepsilon_j^2\right)\mathbf{A}^{N-5} - \cdots\right]. \tag{13}$$

Now that we have obtained a power series expansion of the Green's function, we can correlate this expression with graph theoretic thinking. A walk of length $k$ in the graph $\mathbb{G}$ is defined as a set of (not necessarily different) nodes (vertices) $i_1, i_2, \cdots, i_k, i_{k+1}$ such that for all $1 \leq l \leq k, (i_l, i_{l+1}) \in E$. A closed walk is defined as a walk for which $i_1 = i_{k+1}$. It is known that $(\mathbf{A}^k)_{rs}$ counts the number of walks of length $k$ between the nodes $r$ and $s$.[82,83] In a similar way, $(\mathbf{A}^k)_{rr}$ counts the number of closed walks of length $k$ starting (and ending) at the node $r$. The way walks are defined and enumerated is less familiar to chemists, so in the next section we will pause to review the formalism, and in the SI show a brief proof of the relation.

One issue to be addressed is where the coefficients of the power series expansion of the Green's function come from. This is a non-trivial matter, to which we will return.

Continuing with the general development, since $N$ is even (for the cases considered), the exponent of **A** is always odd in eq. 13. This means that the matrix element of the Green's function can be correlated with the number of odd-length walks on the molecular graph. From this finding emerges the conclusion that *if there are only even-length walks between a pair of vertices/atoms i and j, the (i,j) element of the Green's function should be equal to zero.* This zero is precisely the condition for the occurrence of QI between the atoms $i$ and $j$. It should be noted that Estrada had already arrived at the same result by a different approach.[60] Also, the difference in electron transport and QI features between the odd- and even-length transport pathways has been clarified by Pedersen and co-workers, though their classification is based on the number of atoms in the path.[9]

We can connect the above-derived selection rule for the occurrence of QI to the starring procedure of alternant hydrocarbons. In such molecules, the nearest neighbor of a starred atom is always an unstarred atom (see Scheme 1a). So the length of the walks between a starred atom and an unstarred atom is odd. On the contrary, the length of *all* walks between two starred atoms or two unstarred atoms is even. Therefore, QI is expected to occur between a pair of two starred atoms or two unstarred atoms. The same conclusion can be drawn from the SSP approach, as shown by Ernzerhof and co-workers[84] as well as the Sheffield group.[62, 85] Further, this result can alternatively be proved on the basis of the orbital representation of the Green's function proposed by Yoshizawa and coworkers.[86,87,88] The electronic coupling through such a pair of atoms has also been termed "alike" coupling,[89] that between a starred atom and an unstarred atom "disjoint" coupling.[89]

Scheme 2 shows some examples of the expansion of the Green's function in powers of the adjacency matrix for various alternant hydrocarbons. We can see that the sign of the coefficient of $\mathbf{A}^{4n-3}$ is negative, while that of $\mathbf{A}^{4n-1}$ is positive. This is a non-trivial observation; for it implies that, even if there are only odd-length walks between a pair of atoms, it could happen that contributions to the Green's function from the odd-length walks cancel out. This is why we cannot say that the more odd walks between two sites, the greater the transmission probability. When the cancellation is complete, another kind of QI occurs. We will return to this point.

***Scheme 2.*** *Examples of the power series expansion of the Green's function for various alternant hydrocarbons.*

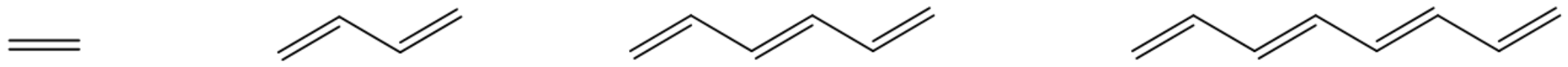

$\mathbf{G} = -\mathbf{A}$  $\mathbf{G} = \mathbf{A}^3 - 3\mathbf{A}$  $\mathbf{G} = -\mathbf{A}^5 + 5\mathbf{A}^3 - 6\mathbf{A}$  $\mathbf{G} = \mathbf{A}^7 - 7\mathbf{A}^5 + 15\mathbf{A}^3 - 10\mathbf{A}$

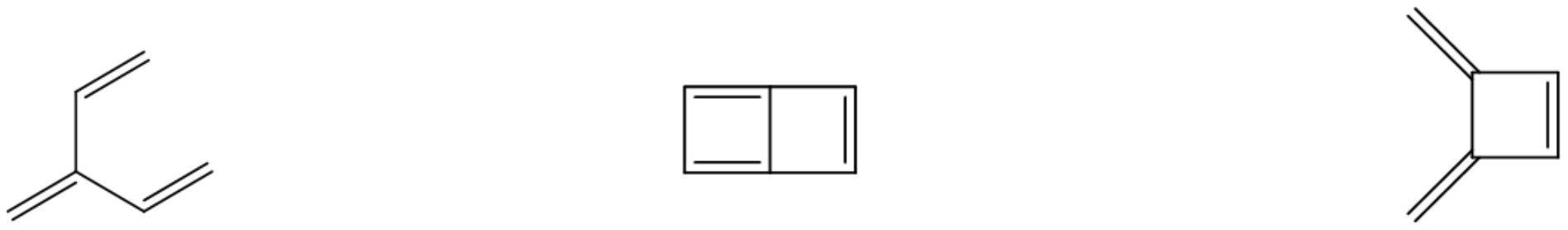

$\mathbf{G} = -\mathbf{A}^5 + 5\mathbf{A}^3 - 5\mathbf{A}$  $\mathbf{G} = -\mathbf{A}^5 + 7\mathbf{A}^3 - 7\mathbf{A}$  $\mathbf{G} = -\mathbf{A}^5 + 6\mathbf{A}^3 - 5\mathbf{A}$

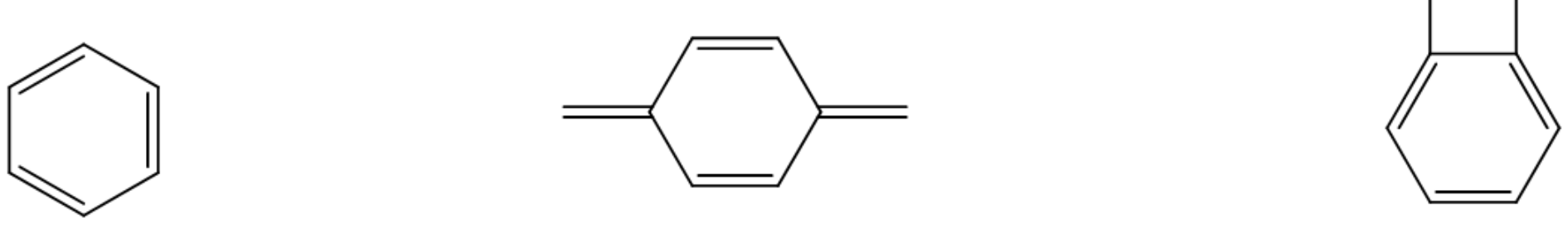

$\mathbf{G} = -\frac{1}{4}\left(\mathbf{A}^5 - 6\mathbf{A}^3 + 9\mathbf{A}\right)$  $\mathbf{G} = \mathbf{A}^7 - 8\mathbf{A}^5 + 18\mathbf{A}^3 - 12\mathbf{A}$  $\mathbf{G} = \mathbf{A}^7 - 9\mathbf{A}^5 + 22\mathbf{A}^3 - 16\mathbf{A}$

### 5.1. An Example, Butadiene

Though the concept of walks is, as we have said, very familiar to the graph-theoretical community, we estimate that it is not so to most chemists. So let's look at an example in some detail, butadiene.

Scheme 3 shows three representations of butadiene, as a Lewis (Kekulé) structure, a molecular graph, and its adjacency matrix. One way to view the entries in adjacency matrix is a representation of the number of walks of length one. There is one such between atoms 1 and 2, 2 and 3, and 3 and 4, ergo the entry 1 for these. The entries of powers of degree $n$ of the adjacency matrix indicate the number of walks of length $n$ between the indices. This is shown graphically for one example, the walks starting from the 1st atom and ending at the 3rd atom as they appear in the (1,3) matrix entry of the second and fourth power of the adjacency matrix (it will become apparent why we have chosen these powers below). The reader can hone his or her understanding here, by calculating the number of walks between other entries.

***Scheme 3.*** *Top: Lewis (Kekulé) structure (left), molecular graph (middle), and adjacency matrix and its inverse (right) for butadiene. Bottom: The 1,3 elements of the second and fourth power of the adjacency matrix, count walks, using 2-steps and 4-steps, respectively, starting from the 1st atom and ending at the 3rd atom. The corresponding matrix element is highlighted by the red dashed circle.*

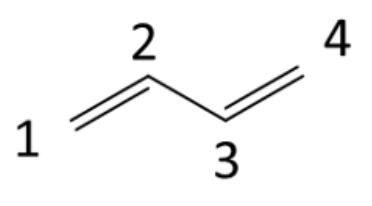

molecule

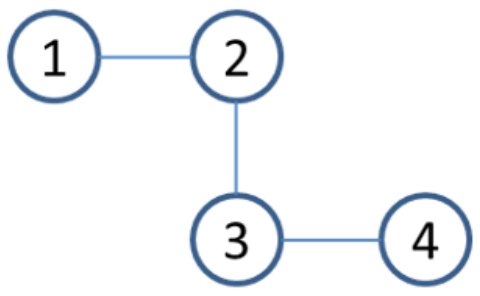

graph

$$\mathbf{A} = \begin{bmatrix} 0 & 1 & 0 & 0 \\ 1 & 0 & 1 & 0 \\ 0 & 1 & 0 & 1 \\ 0 & 0 & 1 & 0 \end{bmatrix}$$

adjacency matrix

$$\mathbf{A}^{-1} = \begin{bmatrix} 0 & 1 & 0 & -1 \\ 1 & 0 & 0 & 0 \\ 0 & 0 & 0 & 1 \\ -1 & 0 & 1 & 0 \end{bmatrix}$$

inverse matrix

1→3 connection

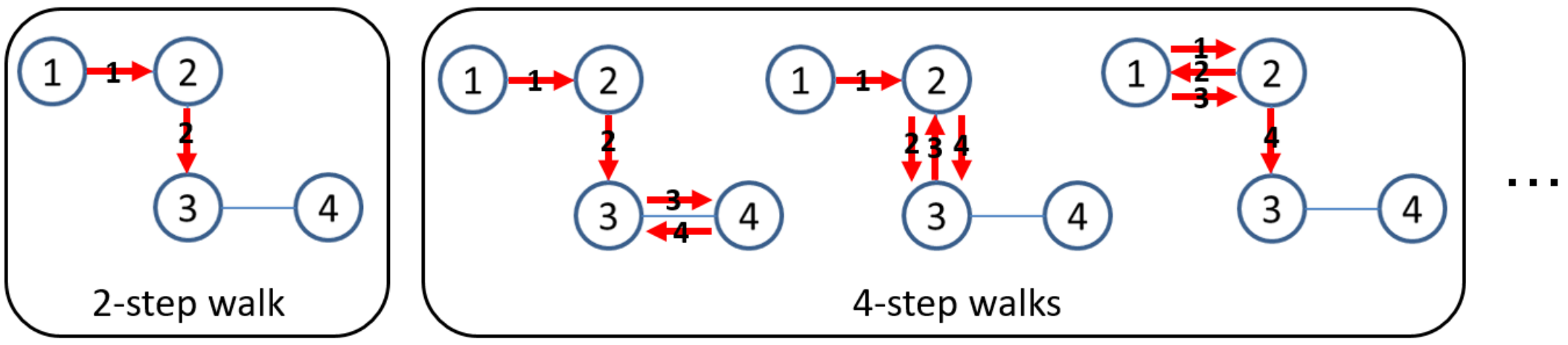

$$\mathbf{A}^{2} = \begin{bmatrix} 1 & 0 & 1 & 0 \\ 0 & 2 & 0 & 1 \\ 1 & 0 & 2 & 0 \\ 0 & 1 & 0 & 1 \end{bmatrix}$$

$$\mathbf{A}^{4} = \begin{bmatrix} 2 & 0 & 3 & 0 \\ 0 & 5 & 0 & 3 \\ 3 & 0 & 5 & 0 \\ 0 & 3 & 0 & 2 \end{bmatrix}$$

As we have shown, QI is associated with zeroes of the Green's function, which is approximated by the negative of the inverse of the adjacency matrix. **G** can be calculated directly, and for a number of π-electron systems, its elements are available in explicit form.[90] Other zeroes of the Green's function are harder to obtain (for what we called hard zeroes,[91] and elsewhere were termed as disjoint cases[89]). Now we do it with walks.

In the previous section, we (and Estrada earlier) showed that $\mathbf{G} = -\mathbf{A}^{-1}$ can be expanded in powers of the adjacency matrix **A**. As eq. 13 shows, only odd powers of **A** appear in that expansion for an alternant hydrocarbon. The elements of odd powers of **A** count the walks of odd number. If all the walks between two specified sites have only an even number of steps, that element must be zero. This is easily confirmed for butadiene, for the 1-3 connection in Scheme 3 – the only walks between these sites have an even number of steps.

But that is not the only way to get QI. *It is also possible for a given matrix element of* $-A^{-1}$ *to vanish (the condition for QI), even when each individual power in its expansion has non-vanishing corresponding matrix elements.* These are the walk equivalent to the hard zero or disjoint QI cases. Let's work through an example for butadiene, the 2,3 connection.

The expansion of **G** in butadiene is symbolically $\mathbf{G} = \mathbf{A}^3 - 3\mathbf{A}$. The explicit matrix elements are shown in Scheme 4, and the specific walks for a (2,3) connection are illustrated in Scheme 5.

***Scheme 4.*** *Power series expansion of the Green's function for butadiene with explicit matrix elements.*

$$\mathbf{G} = \overset{-\mathbf{A}^{-1}}{\begin{bmatrix} 0 & -1 & 0 & 1 \\ -1 & 0 & 0 & 0 \\ 0 & 0 & 0 & -1 \\ 1 & 0 & -1 & 0 \end{bmatrix}} = \underset{\text{3-step walks}}{\overset{\mathbf{A}^{3}}{\begin{bmatrix} 0 & 2 & 0 & 1 \\ 2 & 0 & 3 & 0 \\ 0 & 3 & 0 & 2 \\ 1 & 0 & 2 & 0 \end{bmatrix}}} - 3 \underset{\text{1-step walks}}{\overset{\mathbf{A}}{\begin{bmatrix} 0 & 1 & 0 & 0 \\ 1 & 0 & 1 & 0 \\ 0 & 1 & 0 & 1 \\ 0 & 0 & 1 & 0 \end{bmatrix}}}$$

***Scheme 5.*** *Visualization of the walks on the molecular graph for butadiene of length 3 and 1 starting from the* $2^{nd}$ *atom and ending at the* $3^{rd}$ *atom (top), and that starting from the* $1^{st}$ *atom and ending at the* $2^{nd}$ *atom (bottom).*

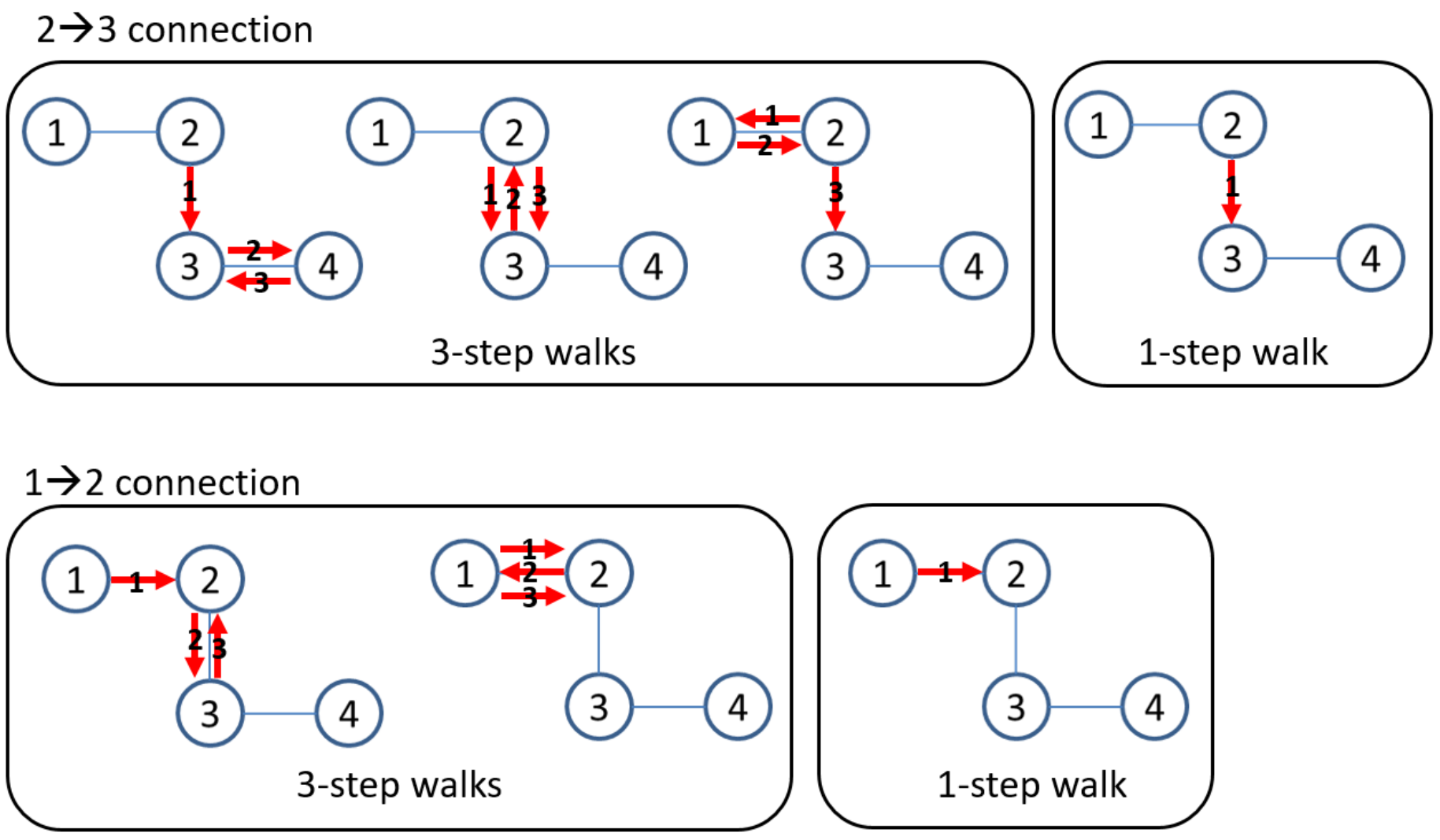

Notice that the (2,3) elements of both **A** and $\mathbf{A}^3$ are not zero – odd-length walks between them are shown explicitly in the Scheme. The (2,3) element of $\mathbf{A}^3$ is 3, which means there are three distinct walks starting from the 2nd atom and ending at the 3rd atom with the length of 3. They are depicted in the top of Scheme 5. The (2,3) element of **A** is 1, which corresponds to a walk of length 1 starting from the 2nd atom and ending at the 3rd atom (see the top of Scheme 5). Since the contribution from the 1-step walk is scaled by a factor of -3, a cancellation occurs between the contributions from the 3-step walks and the 1-step walk. The (2,3) element of **G** is zero.

In contrast to the 2-3 connection, in the case of the 1-2 connection, there are only two walks starting from the 1st atom and ending at the 2nd atom with length of 3, as shown in the bottom of Scheme 5. So the cancellation between the contributions from the 3-step walks and the 1-step walk is incomplete, leading to a finite non-zero matrix element of the Green's function.

In summary, the zero explicit matrix elements in the Green's function for butadiene indicate QI for the corresponding walks. All walks between two starred or two unstarred sites are even in number, thus directly give zeros in this Green's function matrix (easy zeros). Walks between a starred and unstarred site always are odd in number. When zeros for odd walks occur, they are called hard zeros and must be determined by analyzing the explicit matrix elements in the power series expansion of the Green's function for butadiene. The 2-3 connection in butadiene is such a hard zero. Thus, if electrodes are both attached to the 2-3 sites in butadiene dramatic diminution in molecular conductance should be observed.

There is nothing specific to butadiene in the discussion we have just sketched in too much detail. Analogous cancellations due to the algebra of the power expansion can occur for most hydrocarbons (all those drawn in Scheme 2, in particular), and one must watch for them. There are good reasons for their occurrence, as we will see. Here we return to the general discussion.

## 6. NONALTERNANTS

Quantum interference in alternant hydrocarbons is pretty well understood by now. For nonalternants, we think it is fair to say that only partial comprehension is in place; it would be good to have a clear and practical approach for these molecules as well.

What does the power series expansion of the Green's function look like in the case of non-alternants? To this end, we need to return to eq. 5; from there we arrive at

$$p(\mathbf{A}) = (\mathbf{A} - \varepsilon_1\mathbf{I})(\mathbf{A} - \varepsilon_2\mathbf{I})(\mathbf{A} - \varepsilon_3\mathbf{I})\cdots(\mathbf{A} - \varepsilon_N\mathbf{I})$$

$$= \mathbf{A}^N - \left(\sum_i \varepsilon_i\right)\mathbf{A}^{N-1} + \left(\sum_{i,j} \varepsilon_i\varepsilon_j\right)\mathbf{A}^{N-2} - \left(\sum_{i,j,k} \varepsilon_i\varepsilon_j\varepsilon_k\right)\mathbf{A}^{N-3}\cdots + (-1)^N\left(\prod_i \varepsilon_i\right)\mathbf{I}. \tag{14}$$

In a similar way, one can obtain a general expression of the power series expansion of the Green's function for non-alternant hydrocarbons as follows:

$$\mathbf{G} = \frac{(-1)^N}{\prod_i \varepsilon_i}\left[\mathbf{A}^{N-1} - \left(\sum_i \varepsilon_i\right)\mathbf{A}^{N-2} + \left(\sum_{i,j} \varepsilon_i\varepsilon_j\right)\mathbf{A}^{N-3} - \cdots\right]. \tag{15}$$

Note that **A** is not allowed to have a zero eigenvalue (no radicals, no 4*n*-membered rings). Once the eigenvalues of the adjacency matrix are obtained, one can calculate the coefficients in eq. 15. Here one can simplify this equation slightly. If there are no heteroatoms in the molecule, all the diagonal elements of **A** are zero. Since tr**A**=trace of **A** is equal to the sum of **A**'s eigenvalues, namely $\left(\sum_i \varepsilon_i\right) = 0$, tr**A** = 0.[92] Thus, the contribution of the walks with *N*-2 steps must always be zero.

As can be seen in equation 15, generally both odd- and even-power terms of **A** enter for a nonalternant. So it would be difficult to derive from this expression a simple selection rule like the one shown above for alternants. However, one might be able to regard the even-power terms as a perturbation, because the maximum order in the even-power terms is quite generally not so high, and their coefficients are not so large. This will become clear when one applies a scheme to obtain the characteristic polynomial which was proposed by Hosoya, applicable to both alternant and non-alternant hydrocarbons.[93] We will show the workings of Hosoya's scheme below (see section 7.3). Also, as will be discussed in section 11.3, using the binominal theorem, one can derive an infinite power series expansion of the Green's function, where only odd powers of **A** appear, even in the case of non-alternants. Here we begin with a finite power series expansion based on the Cayley-Hamilton theorem.

Let us see some examples of the finite power series expansion of the Green's function for a selection of non-alternant hydrocarbons (see Scheme 6). In triafulvene and fulvene, interestingly we cannot see any even-power term, though one might argue that the last term, the identity matrix, can be viewed as a $0^{th}$-power term of **A**. **I** represents a 0-length walk, which is usually subject to QI in alternant-hydrocarbons, for example, the 1-1 connection of ethylene.[94] One may well call this an even-length walk. But because of the **I** contribution to the expansion, QI is not expected for the 4-4 connection of triafulvene and the 6-6 connection of fulvene. These expectations, a situation very different from ethylene, have already been confirmed in the literature.[16,27]

***Scheme 6.*** *Examples of the power series expansion of the Green's function for non-alternant hydrocarbons: triafulvene, fulvene, [3]radialene, and azulene.*

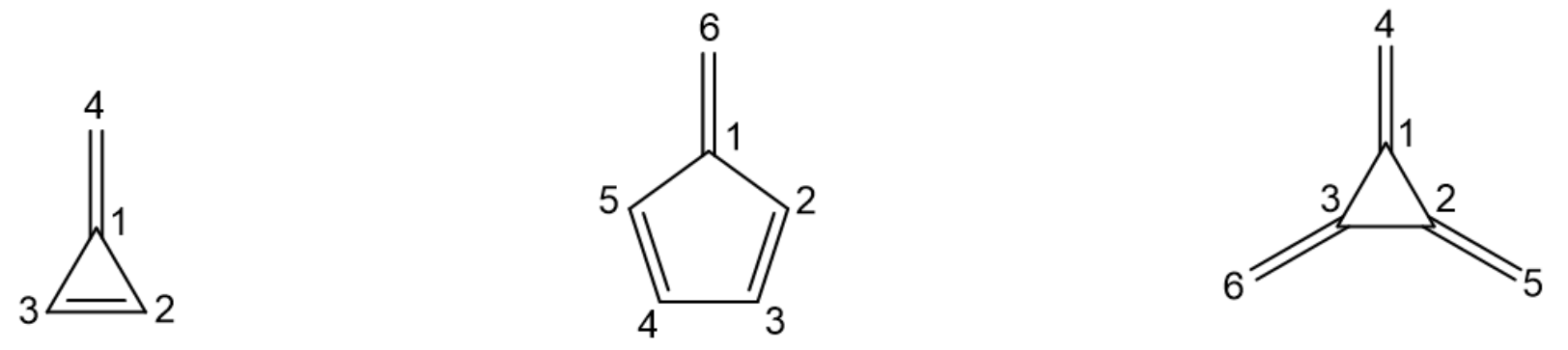

triafulvene — fulvene — [3]radialene

$$\mathbf{G} = \mathbf{A}^3 - 4\mathbf{A} - 2\mathbf{I} \qquad \mathbf{G} = -\mathbf{A}^5 + 6\mathbf{A}^3 - 8\mathbf{A} + 2\mathbf{I} \qquad \mathbf{G} = -\mathbf{A}^5 + 6\mathbf{A}^3 + 2\mathbf{A}^2 - 6\mathbf{A}$$

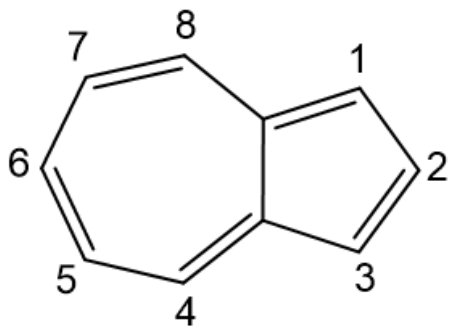

azulene

$$\mathbf{G} = \frac{1}{4}\left[-\mathbf{A}^9 + 11\mathbf{A}^7 - 41\mathbf{A}^5 + 2\mathbf{A}^4 + 61\mathbf{A}^3 - 6\mathbf{A}^2 - 31\mathbf{A} + 2\mathbf{I}\right]$$

No QI features have been observed (calculated) in triafulvene, while the 3-4 and 2-4 connections of fulvene are cases of QI.[27] In these connections one readily realizes that there are odd-length walks between the two electrode attachment sites. Thus, the occurrence of QI in fulvene is not obvious from the power series expansion. The QI must be an outcome of the cancellation of contributions from the odd-length walks.

### 6.1 Fulvene as an Example

Let us actually check the QI feature in fulvene by counting the walks on the molecular graph. One needs to consider 5-, 3-, and 1-step walks because the $5^{th}$, $3^{rd}$, and $1^{st}$ powers of **A** are included. **I** formally represents the 0-step walk, but such a walk does not contribute to the connections of interest. In Scheme 7, we compare the 2-4 and 2-5 connections. Since there is no way to arrive at the C4 or C5 atom from the C2 atom in one step, we only need to consider the 5- and 3-step walks. For the 2-4 connection, the walks proceed in generally anti-clockwise direction, while for 2-5 they are almost clockwise. Both 2-4 and 2-5 connections have a single walk of length 3. As for the walks with the length of 5, the 2-4 connection has one more walk than the 2-5 connection. The difference in the number of the 5-step walks is due to the existence of a branch at the C1 atom, which leads to an extra excursion to the outside of the ring. Since the $\mathbf{A}^3$ term is scaled by 6 in the expansion ($\mathbf{G} = -\mathbf{A}^5+6\mathbf{A}^3-8\mathbf{A}+2\mathbf{I}$), the cancellation is incomplete for the 2-5 connection, leading to a finite transmission probability, while it is complete for the 2-4 connection, resulting in QI. This situation is indeed similar to what we have seen in the case of the 2-3 connection of butadiene.

***Scheme 7.*** *Visualization of the walks (with length of 5 and 3) on the molecular graph for fulvene starting from the 2^nd^ atom and ending at the 4^th^ atom (top) and that starting from the 2^nd^ atom and ending at the 5^th^ atom (bottom).*

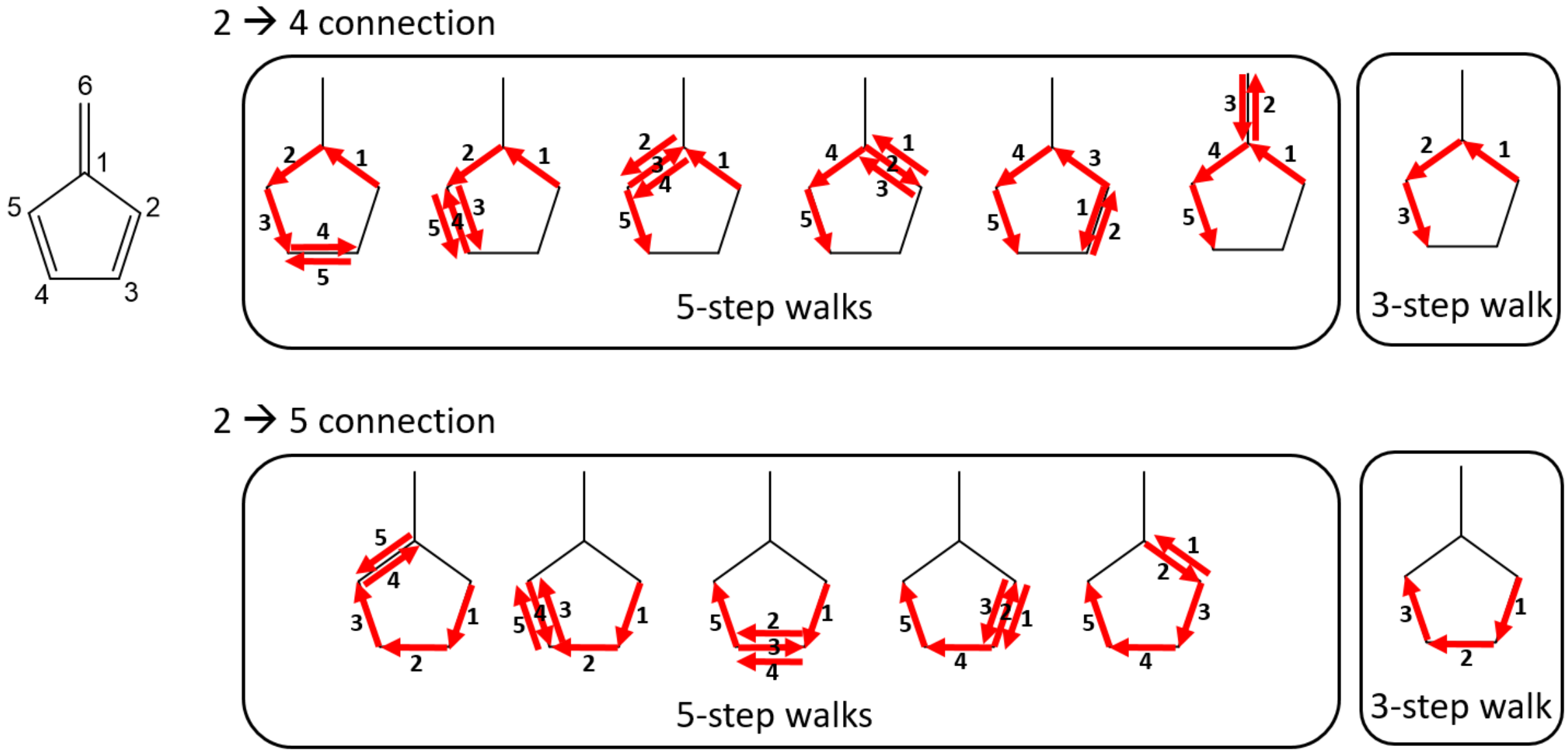

One can attribute the occurrence of QI in the 2-4 connection to the extra branch at the C1 atom, compared to the absence of QI in the 2-5 connection on the basis of walks shown in Scheme 7. Such a branch is regarded as an important QI inducer, namely cross-conjugation.[15,89,95,96] Many cross-conjugated molecules have been theoretically and experimentally found to show a QI feature.

In contrast to triafulvene and fulvene, the power series expansion of the Green's function for [3]radialene and azulene (see Scheme 6 for structure) includes both explicit lower order even-power terms and odd-power terms. But the coefficients of the even-power terms are not high, making them a relatively unimportant perturbation term from the point of view of potential cancellation.

### 6.2 [3]Radialene

Another instructive example is provided by [3]radialene – see Scheme 6 for structure and atom numbering. Since electrodes cannot be attached to the C1, C2 and C3 atoms of [3]radialene, there are only two symmetrically distinct electrode attachment patterns: 4-4 and 4-5. A theoretical calculation predicts that the 4-4 connection leads to QI while the 4-5 connection provides good transmission.[97] We can understand the QI for the 4-4 connection in [3]radialene based on the walks on the molecular graph. Here we need to consider closed walks. It is clear that there are no closed walks of length 1 and length 3 originating from the 4th site, so the closed walks of length 2 and length 5 have to be taken into account. These closed walks are depicted in Scheme 8. There is only one closed walk of length 2 and there are two closed walks of length 5. As can be seen from the power series expansion shown in Scheme 6, the coefficient for $\mathbf{A}^2$ is 2 while that for $\mathbf{A}^5$ is -1. Thus, their contributions cancel out.

***Scheme 8.*** *Top: Closed walk of length 2 and closed walks of length 5 on the molecular graph for [3]radialene originating from the 4th site. Bottom: Walk of length 3 and walks of length 5 starting from the 4th site and ending at the 5th site.*

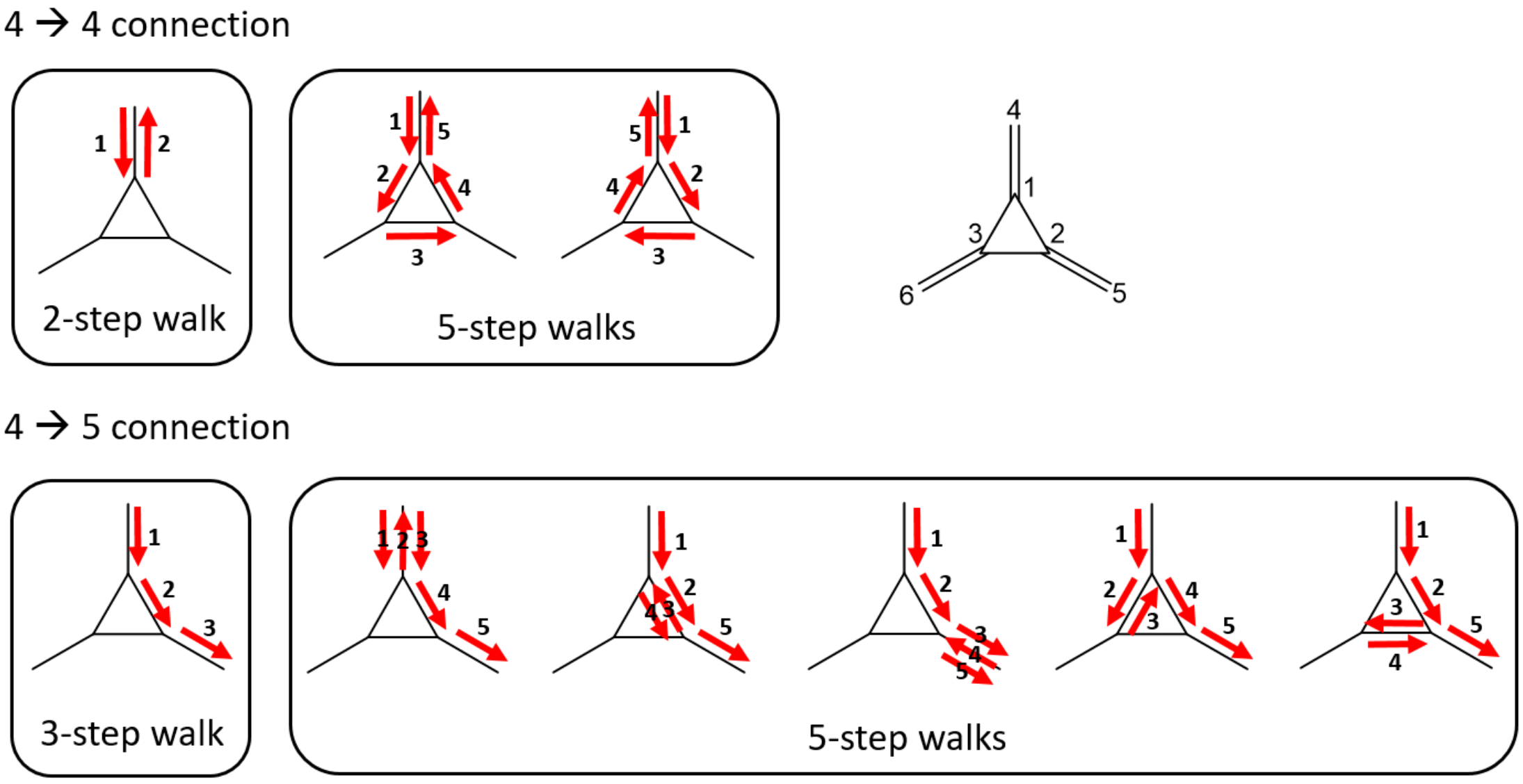

When one considers electron transfer from the 4th site to the 5th site in [3]radialene, no QI is expected. Since the shortest walk between these two sites has a length of 3, walks of length 3 or longer should play an important role in this case. Such walks are depicted in Scheme 8. As can be seen from the power series expansion, the coefficient for $\mathbf{A}^3$ is 6 while that for $\mathbf{A}^5$ is -1. Thus, the cancellation between them is incomplete, leading to a finite electron transmission probability.

In the case of fulvene, the cancellation only occurs between the odd power terms. This is because there are no even power terms in the expansion, except for **I**, whose contribution to transport is often insignificant. However, in the case of [3]radialene, as can be seen above, QI is caused by a cancellation between odd and even power terms.

### 6.3 Azulene, an Important Nonalternant

In azulene, there are some transport pathways which lead to QI, for example, 1-4, 1-6, 1-8, 2-5, and 5-7.[27,97] One of them, namely the 5-7 connection, has been experimentally verified.[98] The explicit matrix elements of the powers of **A** that are involved are shown in SI. For terms higher than the 5th order, all elements are non-zero. This means that if allowed to take a walk with a length longer than 5, one can reach any site on the graph from any other site. The QI features in azulene come from mutual cancellation between the different order terms, including even-order terms. Note that walks of length 2 are not necessarily involved in the QI because walks between some atomic pairs that lead to QI have a length longer than 2, while walks of length 4 are always involved in the QI (see the corresponding matrix elements of $\mathbf{A}^4$ for azulene shown in Scheme S2 in SI). In this respect, the QI features in azulene are akin to those in [3]radialene.

As we have seen in a number of instances, if there is no odd-length walk between a pair of atoms in alternants, QI will occur. But even if there is an odd-length walk between them, QI might still take place. The former QI is what we call "easy-zero" QI, while the latter is what we call "hard-zero" QI.[27,91] The same is true even for QIs in some non-alternant hydrocarbons, whose the Green's function expansion does not include any even power terms that contribute to transport. However, since the finite Green's function expansion for non-alternants includes generally both even and odd powers, a situation may come about where QI is caused by a cancellation between odd and even power terms. Thus, QIs in non-alternants can, in general, fall into the class of the hard-zero QI.

Since a cancellation leading to the hard-zero QI relies on a delicate balance between odd-power terms, the hard-zero QI may not be a robust electronic feature and might be vulnerable to perturbation. Though this is not the place to discuss the matter, this is not a merely nominal distinction, as higher-order effects such as through-space coupling, non-nearest neighbor interactions, and many-body charge-charge correlation make a difference in the QI feature observed between the hard-zero and easy-zero QIs.[9,99]

## 7. WHERE DO THE COEFFICIENTS IN THE POWER EXPANSION COME FROM?

### 7.1. Sachs Graphs

The characteristic polynomial of a molecular graph has been of importance in chemistry because it has consequences, for example, for the topological theory of aromaticity,[100] stability of hydrocarbons,[101] and random walks on molecules (lattices).[102,103] And, as we have seen, that characteristic polynomial, and the Green's function derived from it determine (the interpretation provided by walks on a graph) when QI occurs in a conjugated system, and when it is absent. We need to trace the origin of the coefficients of the power series expansion of the Green's function.

Let us limit our focus initially to alternant hydrocarbons, for simplicity. Under this constraint, eq. 13 can be further simplified. The determinant of a matrix **A** is equal to the product of its all eigenvalues,

$$\det(\mathbf{A}) = \prod_{i}^{N} \varepsilon_i . \tag{16}$$

Since the pairing theorem holds true, eq. 16 can be written as

$$\det(\mathbf{A}) = \left(-\varepsilon_1^2\right)\left(-\varepsilon_2^2\right)\left(-\varepsilon_3^2\right)\cdots\left(-\varepsilon_{N/2}^2\right) = (-1)^{N/2}\prod_{i}^{N/2}\varepsilon_i^2 . \tag{17}$$

By substituting this equation into eq. 13, we obtain

$$\mathbf{G} = \frac{1}{\det(\mathbf{A})}\left[\mathbf{A}^{N-1} - \left(\sum_{i}^{N/2}\varepsilon_i^2\right)\mathbf{A}^{N-3} + \left(\sum_{i\neq j}^{N/2}\varepsilon_i^2\varepsilon_j^2\right)\mathbf{A}^{N-5} - \cdots\right]. \tag{18}$$

Note that we assume **A** is not singular. The determinant of the adjacency matrix actually has a significant chemical meaning. If the system does not include any 4*n* membered ring, the determinant can be correlated with the number of Kekulé structures, *K*, in the molecule, as follows:[104,105]

$$\det(\mathbf{A}) = (-1)^{N/2}K^2 . \tag{19}$$

So det(**A**) in eq. 18 can be replaced with this equation, resulting in

$$\mathbf{G} = \frac{(-1)^{N/2}}{K^2}\left[\mathbf{A}^{N-1} - \left(\sum_{i}^{N/2}\varepsilon_i^2\right)\mathbf{A}^{N-3} + \left(\sum_{i\neq j}^{N/2}\varepsilon_i^2\varepsilon_j^2\right)\mathbf{A}^{N-5} - \cdots\right]. \tag{20}$$

This is why the prefactor of the Green's function for benzene ($K = 2$) is -1/4 (see Scheme 2).[106]

The coefficients of the characteristic polynomial can be obtained solely from the topological structure of the graph, based on a formalism due to Sachs.[105] The process involves a so-called Sachs graph. It should be noted that Coulson presented another scheme to obtain the coefficients in a diagrammatical way.[107] Also, a few different graphical approaches to the polynomial have been explored in the literature.[93,108] Here we follow Sachs's scheme.

Eq. 11, the expansion in powers of **A**, may be rewritten as follows:

$$p(\mathbf{A}) = \sum_{n=0}^{N/2-1} a_{2n}\mathbf{A}^{N-2n} + (-1)^{N/2}K^2\mathbf{I} = \mathbf{0}. \tag{21}$$

The coefficient $a_{2n}$ is the essential part of the Sachs graph and defined as

$$a_{2n} = \sum_{s\in S_{2n}} (-1)^{c(s)}2^{r(s)}, \tag{22}$$

where $s$ indicates a Sachs graph and $S_{2n}$ is the set of all Sachs graphs with $2n$ vertices/atoms. $c(s)$ and $r(s)$ represent the number of components and the number of ring components, respectively. By definition, $a_0 = 1$. By the same scheme applied to eq. 12, we can derive another expression for the Green's function as follows:

$$\mathbf{G} = \frac{(-1)^{N/2}}{K^2}\left[\sum_{n=0}^{N/2-1} a_{2n}\mathbf{A}^{N-2n-1}\right]. \tag{23}$$

So how does one obtain the coefficients $a_{2n}$? In the literature, one can find good instructions for drawing and counting Sachs graphs.[109] Here, in the pedagogical spirit of our review, we work through an example, returning to butadiene.

In the power series expansion of the Green's function for butadiene shown in Scheme 2, the important coefficient of the second term, the one that allowed the cancellation for the 2,3-conenction, is -3. This corresponds to $a_2$. To obtain it, one needs to draw a set of $S_2$ Sachs graphs for butadiene. There are three major strategies for drawing a set of $S_{2n}$ Sachs graphs: 1) find $n$ pairs of non-contiguous bonds, 2) find a cycle (cycles) so that the total number of vertices/atoms included in the cycle (cycles) is $2n$, and 3) find $m$ pairs of non-contiguous bonds and a cycle (cycles) so that the total number of vertices/atoms included in the cycle (cycles) is $2n$-$2m$. $S_2$ is the set of all Sachs graphs with two vertices/atoms. Since it is impossible to draw a cycle consisting of only two vertices/atoms, one can only follow the first strategy. Thus, three sets of $S_2$ Sachs graphs can be drawn for butadiene as shown in Scheme 9. Each Sachs graph includes only one component, so $c(s) = 1$. Since there is no ring, $r(s) = 0$. The $a_2$ coefficient can be calculated as $a_2 = (-1)^1 2^0 + (-1)^1 2^0 + (-1)^1 2^0 = -3$. Thus, it is easy to see that $|a_2|$ is equal to the number of edges/bonds of the graph.

***Scheme 9.*** *Set of $S_2$ Sachs graphs for butadiene.*

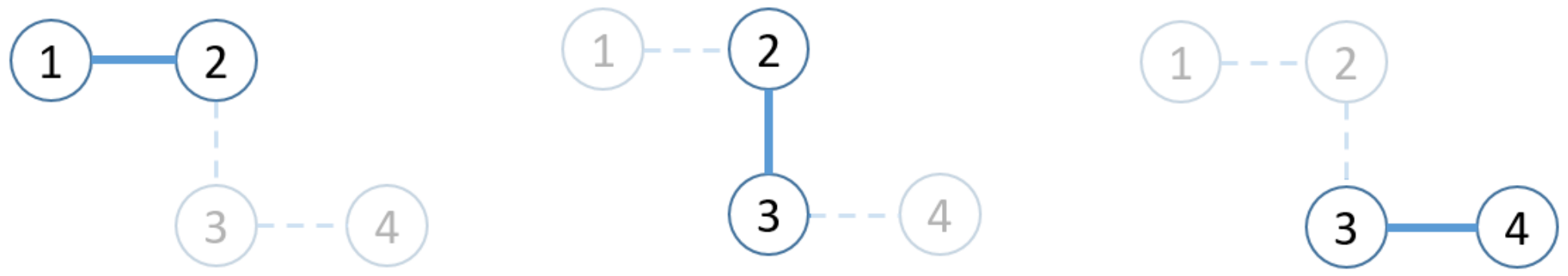

Let us take another example, to get an idea how another coefficient, now $a_4$, may be obtained. This coefficient appears in the characteristic polynomial for a molecule whose number of atoms is 6 or larger. We use dimethylenecyclobutene (see Scheme 10) as an example. Its $a_4$ coefficient is 5 as shown in Scheme 2. Note that there is a prefactor $(-1)^{N/2}$ in eq. 23. To obtain this coefficient in the diagrammatic way, in accordance with the strategies shown above, one can draw a set of $S_4$ Sachs graphs for dimethylenecyclobutene, as shown in Scheme 10. The elements contributing to the summation of eq. 22 are shown under the corresponding Sachs graph. By summing them up, one obtains $a_4 = 5$.

The Sachs graph can provide us with a starting point, a way to think about a conceptual bridge between the topological feature of π-conjugation and QI features on the basis of the graph-theoretic path counting. However, as the size of the molecule gets larger, more effort would be required for enumerating the Sachs graphs. So predicting QI based on the Sachs graph of a complex molecule might not be realistic. Even calculation by brute force from the eigenvalues (see eq. 20) would probably be preferred in case of moderate-sized molecules. For large molecules, it could be that neither the Sachs graph nor the brute force approach of calculating the eigenvalues is in order. This is because our simple Green's function cannot describe the effect of decoherence or loss of electron phase coherence caused by incoherent scattering processes, which cannot be neglected in long-distance electron transport[1,2,110] (see SI for detail).

***Scheme 10.*** *Set of $S_4$ Sachs graphs for dimethylenecyclobutene. The elements contributing to the summation shown by eq. 22 are shown under the corresponding Sachs graph.*

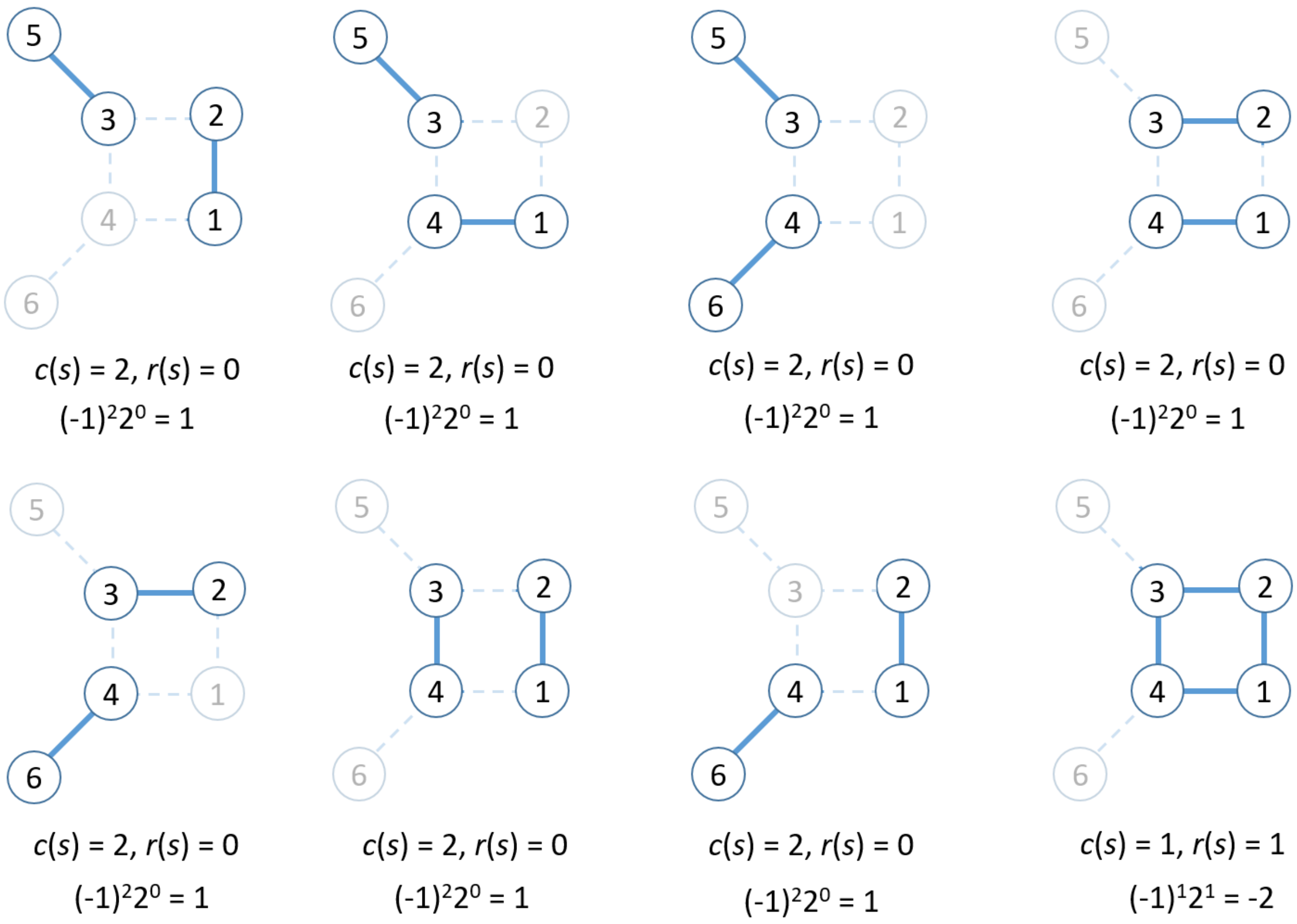

### 7.2. Newton's Identities

In the last section we traced the origin of the expansion coefficients of the Green's function back to Sachs graphs. In this section, we will show an alternative way to obtain the coefficients using Newton's identities.[111] To this end, we again begin with eq. 14, the most general expression for the characteristic polynomial in this manuscript:

$$
\begin{aligned}
p(\mathbf{A}) &= \mathbf{A}^N - p_1\mathbf{A}^{N-1} + p_2\mathbf{A}^{N-2} - p_3\mathbf{A}^{N-3}\cdots + (-1)^N p_N\mathbf{I} \\
&= \sum_{n=0}^{N-1}(-1)^n p_n\mathbf{A}^{N-n} + (-1)^N p_N\mathbf{I} = 0,
\end{aligned} \tag{24}
$$

where $p_0 = 1$ by definition. Also, note that $p_1 = 0$ because $-p_1 = \sum \varepsilon_i = \mathrm{tr}\mathbf{A} = 0$, as already explained around eq. 15. Then (just as we did around eq. 11) another expression for the Green's function can be derived as follows:

$$
\mathbf{G} = \frac{1}{p_N}\sum_{n=0}^{N-1}(-1)^{N-n} p_n\mathbf{A}^{N-n-1}. \tag{25}
$$

To construct the series expansion of the Green's function, one needs to obtain the coefficients of the characteristic polynomial $p_n$. They are determined through Newton's identities, in which the coefficients can be correlated with the trace of $\mathbf{A}^k$ as shown below.[112]

$$
\mathrm{tr}\mathbf{A} + p_1 = 0, \tag{26}
$$

$$
\mathrm{tr}\mathbf{A}^2 + p_1\mathrm{tr}\mathbf{A} + 2p_2 = 0, \tag{27}
$$

$$
\mathrm{tr}\mathbf{A}^3 + p_1\mathrm{tr}\mathbf{A}^2 + p_2\mathrm{tr}\mathbf{A} + 3p_3 = 0, \tag{28}
$$

$$
\mathrm{tr}\mathbf{A}^4 + p_1\mathrm{tr}\mathbf{A}^3 + p_2\mathrm{tr}\mathbf{A}^2 + p_3\mathrm{tr}\mathbf{A} + 4p_4 = 0, \tag{29}
$$

and so on.

The trace of $\mathbf{A}^k$ can be related to the $k^{\mathrm{th}}$ moment[56] of given eigenspectrum $\{\varepsilon_i\}$, which is defined as

$$
\mu_k = \sum_i^N \varepsilon_i^k = \mathrm{tr}\mathbf{H}^k. \tag{30}
$$

There is a body of work, going back to 1967, originating in the paper of Cyrot-Lackmann,[113] which relates walks on a molecule/lattice to moments of energy levels or densities of states. The $k^{\mathrm{th}}$ moment has a graph-theoretic meaning of closed $k$-step walks through the graph. Structural trends, relative energies as a function of electron count can be related directly to geometry and topology,[114,115] by using a unitary transformation, $\sum \varepsilon_i^k = \mathrm{tr}\mathbf{H}^k$. If we apply eq. 1 to this equation, one can obtain $\mu_k = \beta^k\mathrm{tr}\mathbf{A}^k$. Since we use $\beta$ as the unit of energy, we may omit it from the equation. The $0^{\mathrm{th}}$ moment $\mu_0$ measures the total number of states, the $1^{\mathrm{st}}$ moment $\mu_1$ the center of gravity of the eigenspectrum, the $2^{\mathrm{nd}}$ moment $\mu_2$ describes the mean square width of the eigenspectrum, the $3^{\mathrm{rd}}$ moment $\mu_3$ its skewness, and the $4^{\mathrm{th}}$ moment $\mu_4$ gives a measure of unimodal versus bimodal behavior in the eigenspectrum.[114]

Given a sufficient number of moments, the energy distribution of the molecules/solid can be recovered. And through the relationship of moments with the powers of the adjacency matrix, therefore the walks in the molecule, a clear relationship of energy to atom connectivity emerges.

Using the moments, we can rewrite the equations 26-29 as, respectively:

$$\mu_1 + p_1 = 0, \tag{31}$$

$$\mu_2 + p_1\mu_1 + 2p_2 = 0, \tag{32}$$

$$\mu_3 + p_1\mu_2 + p_2\mu_1 + 3p_3 = 0, \tag{33}$$

$$\mu_4 + p_1\mu_3 + p_2\mu_2 + p_3\mu_1 + 4p_4 = 0. \tag{34}$$

These equations look very simple, and one needs to notice that $p_1 = -\mu_1 = 0$. By solving these equations sequentially, one can obtain expressions for the coefficients of the characteristic polynomial as, respectively: [112,116]

$$p_1 = -\mu_1, \tag{35}$$

$$p_2 = -\frac{1}{2}\mu_2, \tag{36}$$

$$p_3 = \frac{1}{3}\mu_3, \tag{37}$$

$$p_4 = -\frac{1}{4}\mu_4 + \frac{1}{8}\mu_2^2. \tag{38}$$

These equations are of importance because they imply $p_n$ is a function of the $k^{\text{th}}$ moment, where $k$ must not exceed $n$. The coefficients in equations 35-38 can be determined through Young's diagram;[112] we are working on their physical/chemical meaning.

### 7.3. Non-Adjacent Numbers

There is another approach to the physical/chemical interpretation of the coefficients which appear in the power expansion of the Green's function. This is the "non-adjacent number" concept, a name proposed by Hosoya in his 1971 seminal paper,[117] which gave birth to the study of discrete topology in chemistry. Nowadays, this number is known in mathematics as the "edge independence number" or "matching number" and the polynomial related to it is known as the "matching polynomial".[118] In this paper, we will use Hosoya's original terminology. The non-adjacent number $q(\mathbb{G}, k)$ can be defined as the number of ways for choosing $k$ non-contiguous edges/bonds in a graph $\mathbb{G}$ or a molecule. One can use $q(\mathbb{G}, k)$ to write down the characteristic polynomial for tree graphs $\mathbb{G}$ or acyclic $\pi$-conjugated hydrocarbons as[93,117]

$$p(\lambda) = \sum_{k=0}^{N/2} (-1)^k q(\mathbb{G},k)\lambda^{N-2k} \quad . \tag{39}$$

By applying the Cayley-Hamilton theorem, one can convert this expression into the power series expansion of the Green's function with the same technique that we have applied in several places above.

$$p(\mathbf{A}) = \sum_{k=0}^{\frac{N}{2}-1} (-1)^k q(\mathbb{G},k)\mathbf{A}^{N-2k} + (-1)^{N/2} q(\mathbb{G},N/2)\mathbf{I} = 0$$

$$\mathbf{G} = \frac{1}{q(\mathbb{G},N/2)} \sum_{k=0}^{\frac{N}{2}-1} (-1)^{k+\frac{N}{2}} q(\mathbb{G},k)\mathbf{A}^{N-2k-1} \quad . \tag{40}$$

One can find that the power of **A** is an odd number when *N* is even. This is because all tree graphs/acyclic π-conjugated hydrocarbons are bipartite graph/alternant hydrocarbons.

In Scheme 11 we illustrate how one can derive the non-adjacent number, taking hexatriene and vinylbutadiene as an example. Here hexatriene is a prototypical example for a linear polyene, while vinylbutadiene is used as that for a branched polyene. The application of $q(\mathbb{G},k)$ to hexatriene is already in the literature,[26] but for pedagogical clarity we reproduce it here. $q(\mathbb{G},0)$ means the number of ways in which one does not choose any edges/bonds. This is always 1. As for $q(\mathbb{G},k)$, where $k \geq 1$, we highlight non-contiguous edges/bonds by a bold line. $q(\mathbb{G},1)$ must coincide with the number of the edges/bonds. This is akin to the set of $S_2$ Sachs graphs, as shown in Scheme 9. One can clearly see a correspondence between $q(\mathbb{G},k)$, the coefficients in the characteristic polynomial, and the Green's function expansion. The coefficient of even-power terms in the characteristic polynomial leads to that of odd-power terms of the Green's function whose power is one less than that of the corresponding even-power term of the characteristic polynomial.

***Scheme 11.*** *Illustration of counting the non-adjacent number* $q(k)$ *for hexatriene (a) and vinylbutadiene (b). Chosen non-contiguous edges/bonds are highlighted by a bold line. Next to the structure of the molecules, the characteristic polynomial and the power series expansion of the Green's function are shown.*

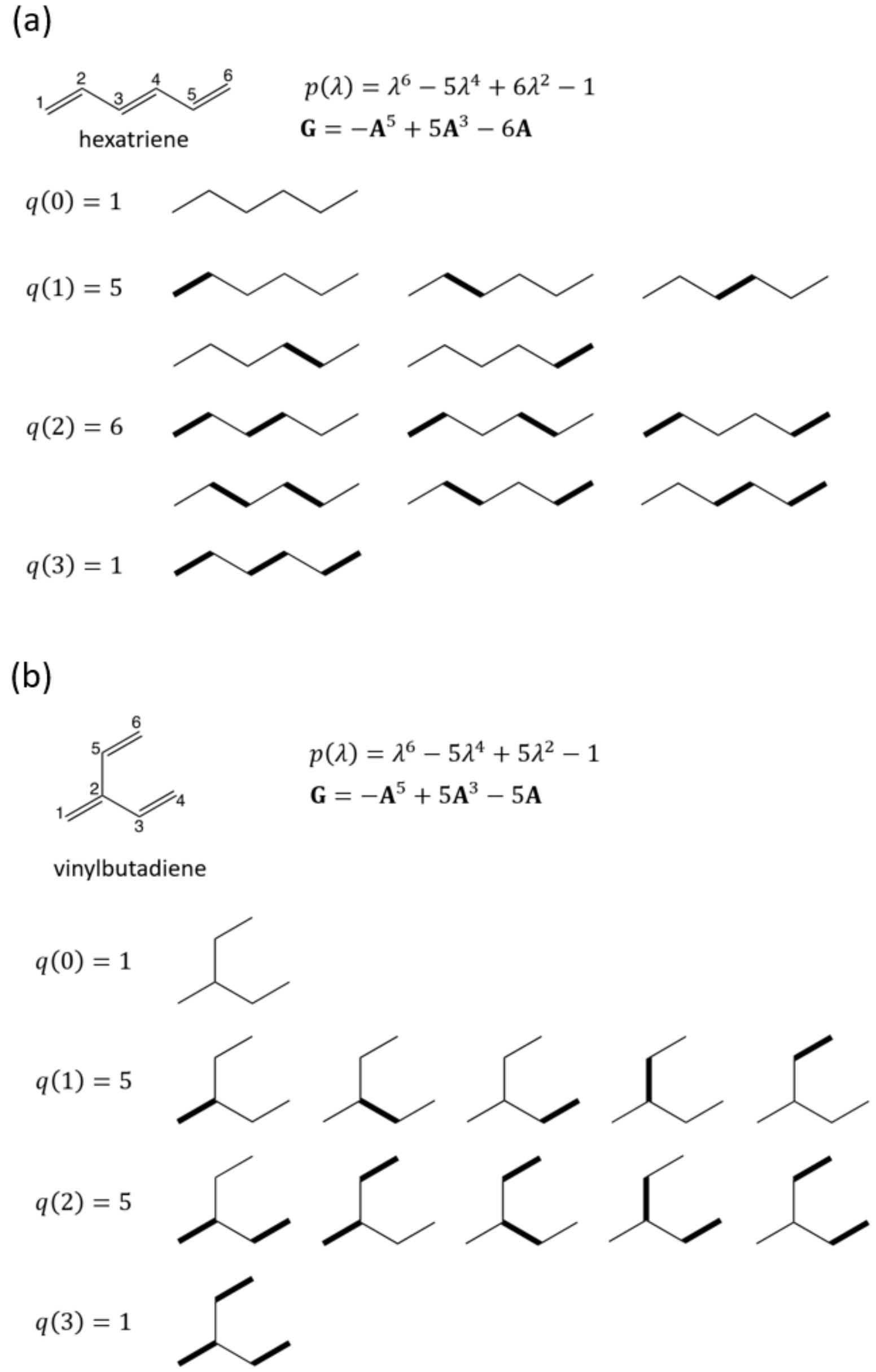

One might be tempted to replace the bold lines in Scheme 11 with double bonds. An illustration of such a drawing is shown in Scheme 12. Now one can speculate that there should be a correlation between the coefficients and the number of "covalent" valence-bond (VB) structures, excluding ionic structures. As can be seen from the scheme, $q(\mathbb{G}, N/2)$ corresponds to the number of closed-shell classical structures, or one might call it the number of all-bonded Lewis or Kekulé resonance structures. Such a structure should be a major contributor to the VB description of π-systems. It has been recently demonstrated that the importance of such a structure actually gets smaller as the π-conjugation gets longer in linear polyenes.[119] $q(\mathbb{G}, N/2-1)$, which is the absolute value of the coefficient of $\lambda^2$, can be interpreted as the number of the possible diradical structures, while $q(\mathbb{G}, N/2-2)$, which is the absolute value of the coefficient of $\lambda^4$, can be interpreted as the number of the possible tetraradical structures. Some of them are Rumer structures,[120,121] but most of them are not. So they might not be important in a practical VB calculation.

***Scheme 12.*** *Illustration of counting the VB structures* $q(k)$ *for hexatriene (a) and vinylbutadiene (b). Chosen non-contiguous edges/bonds are indicated by double bonds. The remaining vertices/carbon atoms are depicted as radical centers. Next to the structure of the molecules, the characteristic polynomial and the power series expansion of the Green's function are shown.*

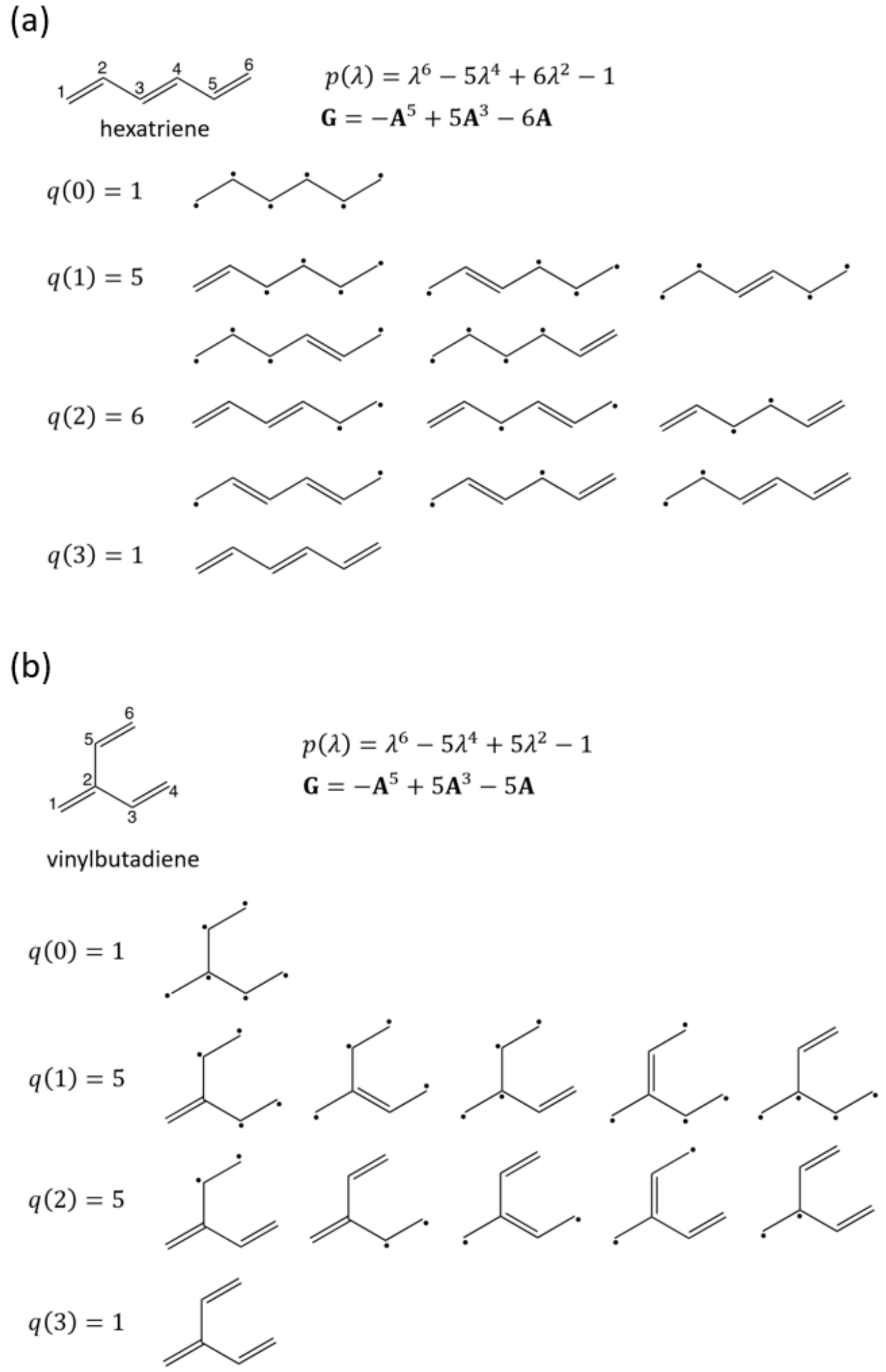

Eventually one arrives at $q(\mathbb{G}, 0)$, which corresponds to the structures in which all double bonds are broken and so all carbon atoms bear an unpaired electron. Since there exists only one such structure, $q(\mathbb{G}, 0) = 1$.

We have pointed elsewhere at a potential correlation between molecular conductance, especially quantum interference, and diradical character.[27,77,122] Here, again, we have a feeling that an underlying theory should connect the number of diradical VB structures and the coefficients of the Green's function expansion. And this we intend to pursue in the future.

Hosoya has extended the concept of the non-adjacent number to non-tree graphs, including alternant hydrocarbons with rings and non-alternant hyrdrocarbons.[26,93,123] The extended formula reads

$$p(\lambda) = \sum_{k=0}^{N/2} (-1)^k q(k, \mathbb{G}) \lambda^{N-2k} + \sum_i (-2)^{r_i} \sum_{k=0}^{(N-n_i)/2} q\left(k, \mathbb{G} \ominus \mathbb{C}^i\right) \lambda^{N-n_i-2k} \quad , (41)$$

where $\mathbb{G} \ominus \mathbb{C}^i$ indicates a subgraph of $\mathbb{G}$ obtained by deleting a cycle or cycles $\mathbb{C}^i$ and all the edges connecting to $\mathbb{C}^i$, $r_i$ is the number of line-disjoint cycles deleted, and $n_i$ is the number of vertices in $\mathbb{C}^i$. The first term of the right hand side of eq. 41 is the same as eq. 39 and the second term is a correction term due to the presence of a cycle. One may call it a "cycle (ring) correction."

In the case of alternant-hydrocarbons with cycles, every cycle is an even-membered ring, so $n_i$ is always even, making $N - n_i - 2k$ an even number. Therefore, the power of the cycle correction terms should also be even. No odd-power terms emerge from eq. 41. By contrast, in the case of non-alternant hydrocarbons, by definition there are odd-membered cycles in the structure. So $n_i$ takes an odd number, leading to the emergence of odd-power terms in the cycle correction. Since the odd-power terms in the characteristic polynomial eventually lead to the even-power terms in the Green's function expansion in powers of the adjacency matrix, the coefficients of the even-power terms in the Green's function expansion can be traced back to the non-adjacent number or the number of radical VB structures for a subgraph in which odd-membered cycles are deleted.

Since the coefficient of $\lambda^n$ in a characteristic polynomial leads to the coefficient of $\mathbf{A}^{n-1}$ in the Green's function expansion, the coefficient of $\lambda^{N-n_i-2k}$ in the cycle correction can be correlated with the coefficient of $\mathbf{A}^{N-n_i-2k-1}$, where $N - n_i - 2k - 1$ is an even number. Since $n_i$ takes on minimally the value 3 (triangle, the smallest cycle), the degree of the even-power term of the Green's function expansion is *N*-4 or less. For example, the highest even-power term in the Green's function expansion for azulene is $\mathbf{A}^4$ (see Scheme 6). This is because $N - n_i - 2k - 1 = 4$, where $N = 10$, $k = 0$, and $n_i = 5$. Note that the smallest cycle in azulene is the pentagon. The higher order terms should only be of odd powers, even for non-alternant hydrocarbons.

Here we would like to show how the Green's function expansion can be obtained from eq. 41 by taking bicyclo[3.1.0]hexatriene as an example. Scheme 13 shows the structure of bicyclo[3.1.0]hexatriene and its graph $\mathbb{G}$. Three cycles are included in $\mathbb{G}$, namely the triangle $\mathbb{C}^1$, pentagon $\mathbb{C}^2$, and hexagon $\mathbb{C}^3$. The subgraphs of $\mathbb{G}$ obtained by deleting these cycle graphs are also shown. When one eliminates the pentagon $\mathbb{C}^2$ from $\mathbb{G}$ with edges incident to it, the remaining graph is just a vertex corresponding to the C1 atom, which is a null graph $\mathbb{K}_1$. We draw it as a dot. When one eliminates the hexagon $\mathbb{C}^3$ from $\mathbb{G}$ with edges incident to it, there is no remaining graph, which is another null graph $\mathbb{K}_0$. We express it as $\emptyset$.

***Scheme 13.*** *Structure of bicyclo[3.1.0]hexatriene and its graph* $\mathbb{G}$ *(top), the cycle graphs included in* $\mathbb{G}$ *(middle), and subgraphs of* $\mathbb{G}$ *obtained by deleting the cycle graphs (bottom).*

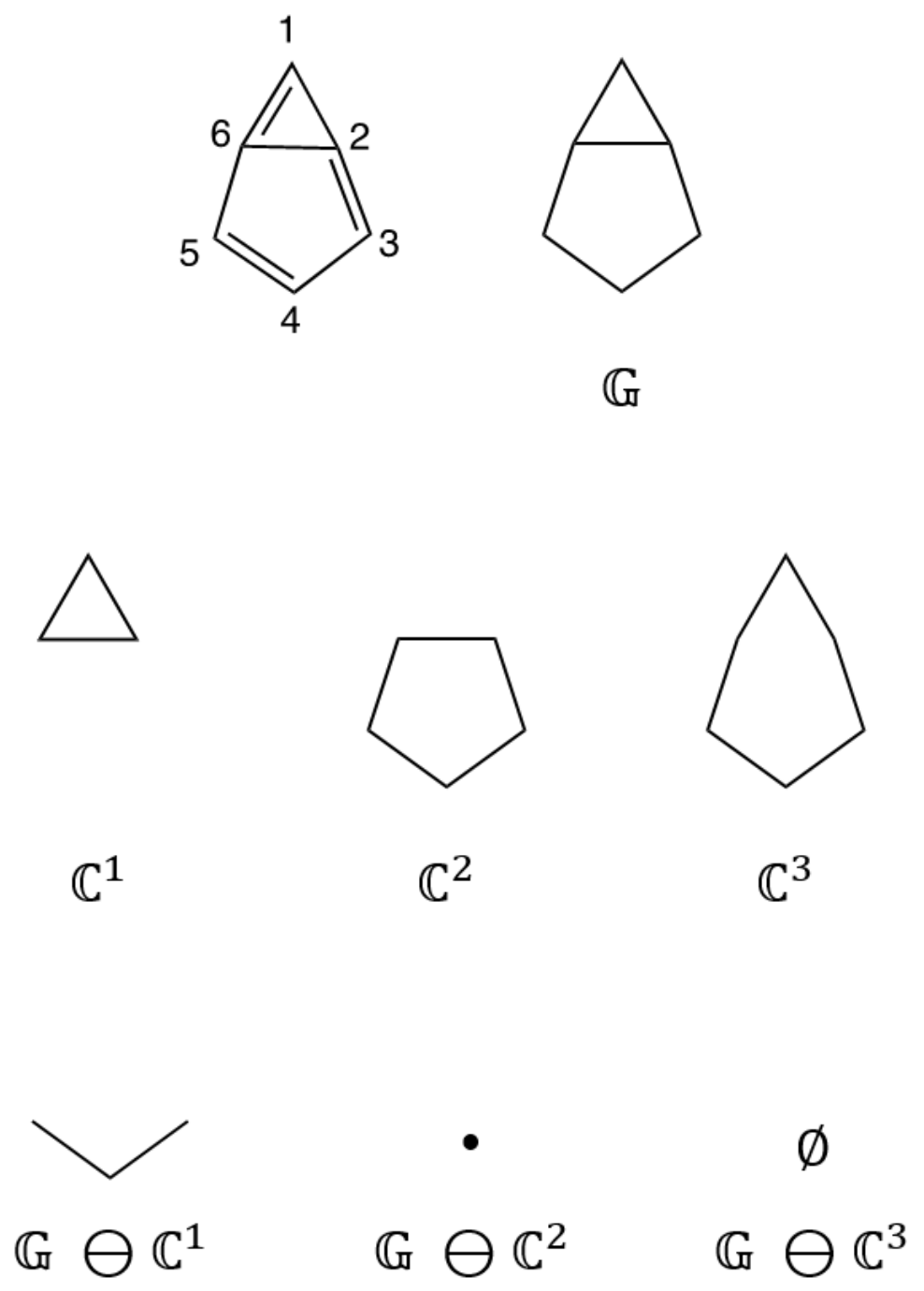

The application of eq. 41 to bicyclo[3.1.0]hexatriene results in $p = p_0 + p_1 + p_2 + p_3$, where $p_0$ is a component of the characteristic polynomial coming from the graph $\mathbb{G}$, or the so-called matching polynomial for the graph, and $p_1$, $p_2$, and $p_3$ are the cycle correction terms corresponding to the subgraphs, $\mathbb{G} \ominus \mathbb{C}^1$, $\mathbb{G} \ominus \mathbb{C}^2$, and $\mathbb{G} \ominus \mathbb{C}^3$, respectively. Scheme 14 illustrates the counting of the VB structures for the graph $\mathbb{G}$ and its subgraphs. Since the total number of vertices included in $\mathbb{G}$ is an even number, the VB structures for $\mathbb{G}$ possess even-numbered radical centers. $p_0$ has only even-ordered terms of $\lambda$, which are converted into odd-ordered terms of $\mathbf{A}$ in the Green's function expansion. As for the subgraphs $\mathbb{G} \ominus \mathbb{C}^1$ and $\mathbb{G} \ominus \mathbb{C}^2$, since an odd-membered cycle is deleted, odd-numbered vertices remain, leading to an enumeration of VB structures with odd-numbered radical centers, such as monoradicals and triradicals. For the case of the subgraph, $\mathbb{G} \ominus \mathbb{C}^3$, which includes nothing (the order-zero graph), $q(0, \mathbb{G} \ominus \mathbb{C}^3) = 1$ by definition.

***Scheme 14.*** *Illustration of counting the VB structures for bicyclo[3.1.0]hexatriene (a) and its*

*cycle-deleted subgraphs (b). Chosen non-contiguous edges/bonds are indicated by double bonds. The remaining vertices/carbon atoms are depicted as radical centers. Next to the structure of the molecules, the components of the characteristic polynomial* $p_0$, $p_1$, $p_2$, *and* $p_3$ *are shown. Note that* $p_0$ *is the matching polynomial for the graph* $\mathbb{G}$.

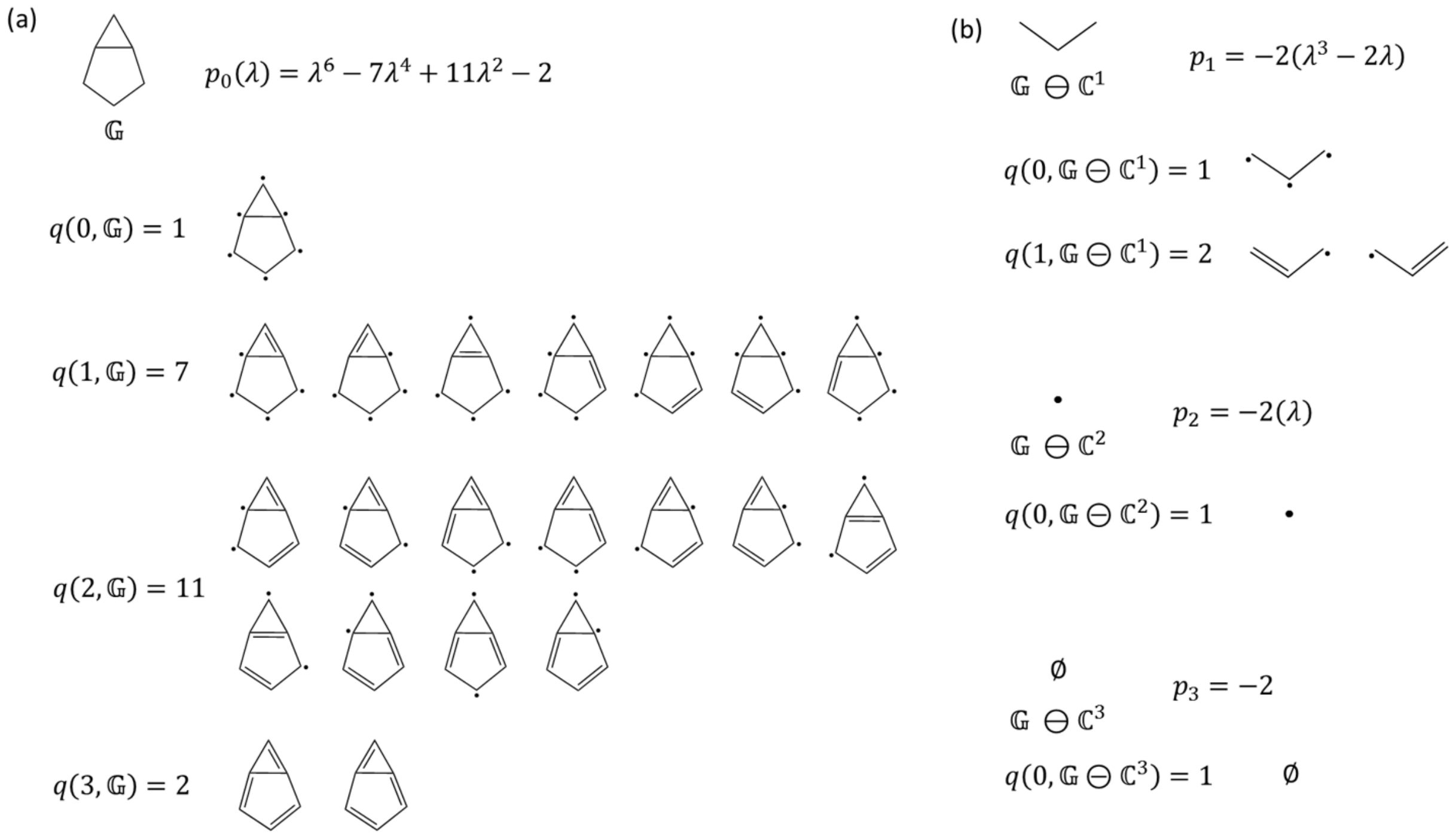

The non-adjacent numbers for the subgraphs shown above need to be scaled by $(-2)^{r_i}$, where $r_i$ is the number of cycles deleted. In the case shown in Scheme 14, $r_i = 1$. If a molecule is so large that one can delete two or more disjoint cycles from the graph, $r_i$ should be more than one. Even in the case of bicyclo[3.1.0]hexatriene, we can see two cycles, a triangle and a pentagon. But they are not a disjoint pair of cycles, for they share an edge.

The matching polynomial for the parent graph of bicyclo[3.1.0]hexatriene is $\lambda^6 - 7\lambda^4 + 11\lambda^2 - 2$ and the polynomial for the cycle correction is $-2\lambda^3 + 2\lambda - 2$. It is essential to note that the polynomial for the parent graph includes only even-power terms while that for the cycle correction potentially includes both the even- and odd-power terms because there is a situation where not only an odd-membered cycle but also an even-membered cycle can be deleted. By combining these two polynomials, we have $p(\lambda) = \lambda^6 - 7\lambda^4 - 2\lambda^3 + 11\lambda^2 + 2\lambda - 4$, leading to $\mathbf{G} = -\frac{1}{4}[\mathbf{A}^5 - 7\mathbf{A}^3 - 2\mathbf{A}^2 + 11\mathbf{A} + 2\mathbf{I}]$. Note the even power term.

## 8. THE CONNECTION TO SCATTERING THEORY AND SOURCE-AND-SINK POTENTIAL APPROACHES

Stadler, Ami, Joachim, and Forshaw,[17] in constructing their visualization scheme for quantum interference, rely on an analytical scattering formalism based on the Hückel matrix. Suppose two electrodes are attached to sites $i$ and $j$ in a molecule; then the transmission coefficient between them can be written as[124]

$$T(i,j) = |S_{ij}|^2, \tag{42}$$

where $S_{ij}$ is the $(i,j)$ entry of the scattering matrix. It has the following form:

$$S_{ij} = \frac{2\Gamma_{ij} i}{\Gamma_{ii}\Gamma_{jj} - \Gamma_{ij}^2 - 1 - (\Gamma_{ii} + \Gamma_{jj})i}, \tag{43}$$

where the matrix element $\mathbf{\Gamma}_{ij}$ is defined as $\det(\mathbf{M}_{ij})/\det(\mathbf{M})$, by using the secular determinant for the adjacency matrix $\mathbf{A}$, namely $\det(\mathbf{M})$, and its $(i,j)$ minor $\det(\mathbf{M}_{ij})$. So one can conclude that when $\det(\mathbf{M}_{ij})$ is zero, the $(i,j)$ element of the scattering matrix should also be zero, resulting in QI. Stadler, Ami, Joachim, and Forshaw found that the terms in the expansion of the minor determinant can be related to a path between the $i^{\text{th}}$ and $j^{\text{th}}$ sites, where all sites of the molecule have to be either traversed within the path or within a closed loop.

In another important application of graph theoretic ideas to molecular conductance, Fowler, Pickup and Todorova[63] utilize the SSP model.[75,76] The transmission probability within the SSP framework reads as

$$T(E) = \frac{4\sin q_L \sin q_R (\tilde{u}\tilde{t} - \tilde{s}\tilde{v})}{\left|e^{-i(q_L + q_R)}\tilde{s} - e^{-iq_R}\tilde{t} - e^{-iq_L}\tilde{u} + \tilde{v}\right|}, \tag{44}$$

where $q_L$ and $q_R$ are the wave vectors of the travelling waves in the left and right one dimensional leads, respectively. And $\tilde{s}, \tilde{t}, \tilde{u},$ and $\tilde{v}$ are defined as, respectively:

$$\tilde{s} = \det[\mathbf{A} - E\mathbf{I}], \tag{45}$$

$$\tilde{t} = \tilde{\beta}_L \det[\mathbf{A} - E\mathbf{I}]^{i,i}, \tag{46}$$

$$\tilde{u} = \tilde{\beta}_R \det[\mathbf{A} - E\mathbf{I}]^{j,j}, \tag{47}$$

$$\tilde{v} = \tilde{\beta}_L \tilde{\beta}_R \det[\mathbf{A} - E\mathbf{I}]^{ij,ij}, \tag{48}$$

where $\tilde{\beta}_L$ ($\tilde{\beta}_R$) is a parameter indicating the coupling strength between the $i^{\text{th}}$ ($j^{\text{th}}$) atom in the molecule and the left (right) lead, and the superscript notation for the determinant is used to indicate the rows and columns struck out of the $N \times N$ $\det[\mathbf{A} - E\mathbf{I}]$. For example, $\det[\mathbf{A}]^{i,j}$ means the determinant of a matrix $\mathbf{A}$ from which the $i^{\text{th}}$ row and $j^{\text{th}}$ column have been removed. Note that $\det[\mathbf{A}]^{i,j}$ is the $(i,j)$ minor of $\mathbf{A}$. By applying the Jacobi/Sylvester determinantal identity,[125,126] we have

$$\tilde{u}\tilde{t} - \tilde{s}\tilde{v} = \tilde{\beta}_L \tilde{\beta}_R \left(\det[\mathbf{A} - E\mathbf{I}]^{i,j}\right)^2. \tag{49}$$

Thus, the SSP model also indicates the $(i, j)$ minor of the secular determinant plays an important role in determining whether QI occurs or not. Fowler, Pickup, Todorova call this minor an opacity polynomial, finding relations between transmission and some chemical concepts.

The studies just cited started from two different origins, yet converged to the same determining factor, namely the $(i,j)$ minor of the secular determinant. It is important to note that the secular determinant is equivalent to the characteristic polynomial, which is the starting point of our work. On multiplication by a factor $(-1)^{i+j}$, the $(i,j)$ minor changes to the $(i,j)$ cofactor. Thus, one can find a way of relating the minor with the characteristic polynomial through the cofactor expansion as follows:

$$p(\lambda) = \det[\mathbf{A} - \lambda\mathbf{I}] = \sum_{k=1} (-1)^{i+k} [\mathbf{A} - \lambda\mathbf{I}]_{ik} \det[\mathbf{A} - \lambda\mathbf{I}]^{i,k}, \tag{50}$$

or

$$p(\lambda) = \det[\mathbf{A} - \lambda\mathbf{I}] = \sum_{k=1} (-1)^{k+j} [\mathbf{A} - \lambda\mathbf{I}]_{kj} \det[\mathbf{A} - \lambda\mathbf{I}]^{k,j}, \tag{51}$$

where $[\mathbf{A} - \lambda\mathbf{I}]_{ij}$ indicates the $(i,j)$ element of the matrix $\mathbf{A} - \lambda\mathbf{I}$. The former corresponds to the cofactor expansion along the $i^{\text{th}}$ row and the latter corresponds to that along the $j^{\text{th}}$ column.

To clarify the correspondence between the Stadler, Ami, Joachim, Forshaw and the Fowler, Pickup, Todorova approaches and ours, we use our beloved example, i.e., butadiene. The secular determinant for butadiene can be written as

$$\det(\mathbf{M}) = \begin{vmatrix} -E & a_{12} & 0 & 0 \\ a_{21} & -E & a_{23} & 0 \\ 0 & a_{32} & -E & a_{34} \\ 0 & 0 & a_{43} & -E \end{vmatrix}, \tag{52}$$

where the off-diagonal elements for adjacent C$i$-C$j$ bonds are represented by $a_{ij}$ instead of using just 1. The reason why we do so will become clear soon. The minors of det(**M**) required for the cofactor expansion along the 2$^{\text{nd}}$ column can be calculated as follows:

$$\det(\mathbf{M}_{12}) = \begin{vmatrix} a_{21} & a_{23} & 0 \\ 0 & -E & a_{34} \\ 0 & a_{43} & -E \end{vmatrix} = a_{12}E^2 - a_{12}a_{34}a_{43}, \tag{53}$$

$$\det(\mathbf{M}_{22}) = \begin{vmatrix} -E & 0 & 0 \\ 0 & -E & a_{34} \\ 0 & a_{43} & -E \end{vmatrix} = -E^3 + a_{34}a_{43}E, \tag{54}$$

$$\det(\mathbf{M}_{32}) = \begin{vmatrix} -E & 0 & 0 \\ a_{21} & a_{23} & 0 \\ 0 & a_{43} & -E \end{vmatrix} = a_{23}E^2, \tag{55}$$

and

$$\det(\mathbf{M}_{42}) = \begin{vmatrix} -E & 0 & 0 \\ a_{21} & a_{23} & 0 \\ 0 & -E & a_{34} \end{vmatrix} = -a_{23}a_{34}E. \tag{56}$$

In the limit of $E \rightarrow 0$, where QI is expected to occur, except for $-a_{12}a_{34}a_{43}$ in det($\mathbf{M}_{12}$), all the other terms will disappear, indicating the occurrence of QI in the 2-2, 2-3, and 2-4 connections. The surviving term can be visualized as shown in Scheme 15, in accordance with the graphical scheme of Stadler et al.[17]

***Scheme 15.*** *Visualization of $a_{12}a_{34}a_{43}$, where an $a_{ij}$ element is described by a black bold line between the $i^{th}$ and $j^{th}$ sites.*

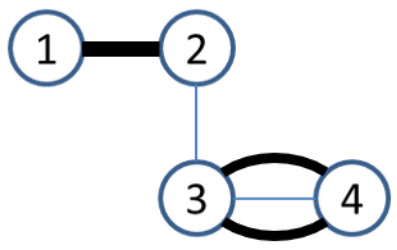

On the basis of eqs. 53-56, one can obtain the expansion of det($\mathbf{M}$), namely the characteristic polynomial of the Hückel matrix for butadiene, as follows:

$$\begin{aligned} \det(\mathbf{M}) &= -a_{12}\det(\mathbf{M}_{12}) - E\det(\mathbf{M}_{22}) - a_{32}\det(\mathbf{M}_{32}) \\ &= E^4 - (a_{12}a_{21} + a_{23}a_{32} + a_{34}a_{43})E^2 + a_{12}a_{21}a_{34}a_{43} \end{aligned}. \tag{57}$$

To obtain an expression for the Green's function, as we have done in section 5, we need to use the Cayley-Hamilton theorem, replacing $E$ with the adjacency matrix $\mathbf{A}$. Eq. 57 can enhance our understanding of Sachs graph and Hosoya's non-adjacent number concept; let us take a look at the visualization of each coefficient shown in Scheme 16. Since $a_{ij} = 1$, $a_{12}a_{21} + a_{23}a_{32} + a_{34}a_{43}$ corresponds to the number of edges and $a_{12}a_{21}a_{34}a_{43}$ corresponds to the number of all-bonded Lewis structures.

***Scheme 16.*** *Visualization of the coefficients which appear in eq. 57 in a similar way to Scheme 15.*

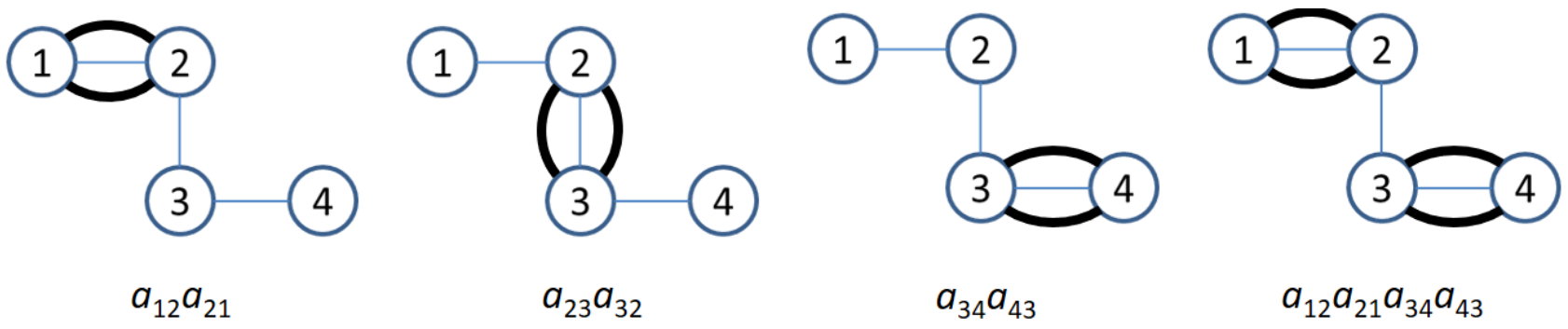

## 9. THE QUADRATIC FORM OF THE EIGENVALUES

We return to the principal equation for the Green's function, eq. 13, which we

reproduce here:

$$\mathbf{G} = \frac{(-1)^{N/2}}{\prod_{i}^{N/2} \varepsilon_i^2} \left[ \mathbf{A}^{N-1} - \left( \sum_{i}^{N/2} \varepsilon_i^2 \right) \mathbf{A}^{N-3} + \left( \sum_{i \neq j}^{N/2} \varepsilon_i^2 \varepsilon_j^2 \right) \mathbf{A}^{N-5} - \cdots \right]. \tag{13}$$

It is possible to make a connection between the quadratic form of the eigenvalues, which one can see in the coefficients of eq. 13, and a concept long of value in chemistry, that of bond indices or bond orders. The product of molecular orbital coefficients can be regarded as a bond order of a certain kind,[127] and has very much to do with the HMO eigenvalues. This is well-documented in the older literature;[65,66] the relationship reads:

$$\varepsilon_i = \sum_{r=1}^{N} C_{ir}^2 \alpha_r + 2 \sum_{(r,s) \in E} C_{ir} C_{is} \beta_{rs}, \tag{58}$$

where $C_{ir}$ is the $i^{\text{th}}$ MO coefficients at atomic site $r$ and the symbol $(r, s) \in E$ indicates that the summation extends only over bonds between those atomic sites or vertices which are linked together in the σ-skeleton, namely the elements of the edges $E$ of the molecular graph $\mathbb{G}$. $\alpha_r$ and $\beta_{rs}$ represent the Coulomb integral of the $r^{\text{th}}$ carbon atom and the resonance integral between the $r^{\text{th}}$ and $s^{\text{th}}$ carbon atoms, respectively. In this paper, they are set to zero and unity, respectively. Therefore, eq. 58 is further simplified as

$$\varepsilon_i = 2 \sum_{(r,s) \in E} C_{ir} C_{is}. \tag{59}$$

By using eq. 59, one of the coefficients of eq. 13 is then

$$\varepsilon_i^2 = 4 \left( \sum_{(r,s) \in E} C_{ir} C_{is} \right)^2 = 4 \sum_{(r,s) \in E} (C_{ir} C_{is})^2 + 8 \sum_{(r,s) \in E} \sum_{(t,u) \in E} (C_{ir} C_{is})(C_{it} C_{iu}). \tag{60}$$

Notice that the last sum in this equation is carried out over all pairs of bonds in the molecule. The first summation is the sum of bond contributions to the corresponding energy $\varepsilon_i$. The second summation accounts for the contribution of the "interaction" of pairs of bonds – not necessarily adjacent – in the molecule.[128] It should be noted that the square of the eigenvalues can be divided into contributions of bonds and those of bond pairs, allowing us to think about the coefficients in terms of the stabilization and destabilization of bonds and pairs of bonds. Obviously, these contributions represent weighted fragment contributions to the energy levels of the molecule, where the fragments are not necessarily connected.

Notice that while the first contribution in eq. 60 – that of bonds – is always positive, the contribution coming from pairs of bonds can be either positive or negative when $i > 1$. The Perron-Frobenius theorem[129] guarantees that the contributions to $\varepsilon_1^2$ are always positive. One can quickly verify this mathematical claim by means of MO theory, which tells us that the lowest energy MO has no node (other than that inherent in being made up of $2p_z$ orbitals), so that all the MO coefficients have the same phase.

Let us go through an example. We apply eq. 60 to butadiene, obtaining

$$\varepsilon_i^2 = 4\left[(C_{i1}C_{i2})^2 + (C_{i2}C_{i3})^2 + (C_{i3}C_{i4})^2\right] + 8\left[(C_{i1}C_{i2})(C_{i2}C_{i3}) + (C_{i1}C_{i2})(C_{i3}C_{i4}) + (C_{i2}C_{i3})(C_{i3}C_{i4})\right]. \tag{61}$$

Apart from the bond contribution given by the first summation, we have the contributions of the fragments C1-C2-C3, C2-C3-C4, and C1-C2 C3-C4 (no bond between C2 and C3). The components of eq. 61, namely bond and pairwise bond contributions for each orbital level, are summarized in Tables 1 and 2. Owing to the pairing theorem, the absolute values of the bond orders and pairwise bond orders for the occupied orbitals are the same as those for the unoccupied counterparts.

***Table 1.*** *Bond contributions to the square of the orbital energies for butadiene. The values in the parentheses are the squared values.*

| $C_rC_s$ | LUMO+1 | LUMO | HOMO | HOMO-1 |
|---|---|---|---|---|
| $C_1C_2$ | -0.224 (0.05) | -0.224 (0.05) | 0.224 (0.05) | 0.224 (0.05) |
| $C_2C_3$ | -0.362 (0.13) | 0.138 (0.02) | -0.138 (0.02) | 0.362 (0.13) |
| $C_3C_4$ | -0.224 (0.05) | -0.224 (0.05) | 0.224 (0.05) | 0.224 (0.05) |

***Table 2.*** *Pairwise bond contributions to the square of the orbital energies for butadiene. Though the squared values of the pairwise bond contributions do not appear in eq. 61, we show the squared values in the parentheses to avoid the sign.*

| $(C_rC_s)(C_tC_u)$ | LUMO+1 | LUMO | HOMO | HOMO-1 |
|---|---|---|---|---|
| $(C_1C_2)(C_2C_3)$ | 0.081 (0.007) | -0.031 (0.001) | -0.031 (0.001) | 0.081 (0.007) |
| $(C_1C_2)(C_3C_4)$ | 0.050 (0.002) | 0.050 (0.002) | 0.050 (0.002) | 0.050 (0.002) |
| $(C_2C_3)(C_3C_4)$ | 0.081 (0.007) | -0.031 (0.001) | -0.031 (0.001) | 0.081 (0.007) |

In Table 1, it is interesting to see that the HOMO and LUMO have the largest amplitudes, or squared values, at the bonds which support the double bonds. And the central bond in butadiene has the lowest amplitude on these orbitals. The same is true for hexatriene (see SI). As for the sign, since the number of nodes in the MO increases with increasing MO energy, the number of negative bond orders also increases. However, in eq. 60, these are squared, so the sign is unimportant.

In terms of the interactions between bonds shown in Table 2, the largest amplitudes, or squared values, in the HOMO and LUMO are obtained for the interaction between the formal double bonds. The same is true for hexatriene (see SI). An observation to be noted in the case of hexatriene is that the interaction of the two closest double bonds (C1-C2 and C3-C4 or C3-C4 and C5-C6) produces a larger amplitude than that between the most distant ones (C1-C2 and C5-C6), leading us to speculate that these pairwise bond interactions might be related to Pauli repulsion between π bonds.[130]

As for the sign of the values in Table 2, we found that the orbitals near the Fermi level, namely HOMO and LUMO, include many pair-wise negative bond orders, while the orbitals far from the Fermi level, namely HOMO-1 and LUMO+1, have no pair-wise negative bond orders. This is because all the bond orders in the HOMO-1 are positive (bonding), leading to positive products, and all the bond orders in LUMO+1 are negative (anti-bonding), also leading to positive products. Therefore, the closer to the Fermi level an orbital lies, the smaller is the summation, i.e., $\sum_{(r,s)\in E}\sum_{(t,u)\in E}(C_{ir}C_{is})(C_{it}C_{iu})$.

It may be possible to derive a correlation between walks and bond interaction (bond order) or pairwise bond interaction (pairwise bond order). To this end, we are working on formulating through-bond and through-space interactions[131] in molecules using graph-theoretic HMO ideas; remote bond-bond interaction has the feeling of through-bond interaction, while pairwise adjacent bond interaction seems more through-space in character.

## 10. HÜCKEL MO PERTURBATION THEORY AND ITS IMPACT ON QI

There are many perturbations of the adjacency matrix that have a direct chemical meaning. For instance, one might change a C atom to another atom (heteroatom substitution).[65,132] Such an effect in a first approximation is only embedded in the diagonal elements; physically it may be seen as an electronegativity perturbation.[133] Other perturbations might change a single off-diagonal entry from zero or one to a certain intermediate value, to express bond formation or dissociation or a conformational change around the bond.[91,134] Still other chemical perturbations might enlarge the matrix, adding atoms. One has to be watchful for perturbations that do not allow a matrix to be inverted.

People in the graph theory community have also been active in addressing perturbation problems because they need to deal with molecular graphs which include heteroatoms.[135] Such molecular graphs can be represented by vertex- and edge-weighted graphs.[100] Dias established a way of obtaining the characteristic polynomial for a molecular graph perturbed by a heteroatom from the characteristic polynomial for the isoconjugate graph (a graph in which the heteroatom is replaced with the carbon atom) and that for the graph obtained upon deletion of heteroatom vertex with its adjacent edges.[136] In this context, the isoconjugate graph corresponds to an unperturbed system.

Recently, Sýkora and Novotný[74] developed an inelastic Hückel model using a graph-theoretical approach, where the electron-vibration coupling is added to a molecular Hamiltonian as a perturbation. Their model allows one to predict inelastic contributions to the conductance due to molecular vibration modes excited by an applied bias voltage.

Before we look at how perturbation affects a power series expansion of the Green's function, we will show an example of just how significantly a small perturbation can change a $\pi$-conjugated system and its characteristic polynomial. There is another reason for what we are about to do – we wish to see if one can understand non-alternant systems (non-bipartite graphs) as derived by perturbation of alternants.

A class of non-alternant hydrocarbons can be generated from an alternant hydrocarbon by forming a bond. Consider [10]annulene, an alternant system. As Scheme 17 shows, by forming one C-C bond (removing two H atoms in the process), one can generate either alternant or non-alternant hydrocarbons. The Hamiltonian/adjacency matrices for the bicyclic compounds are very similar to that for [10]annulene. Hence, one might be able to use a perturbative approach. The adjacency matrices for the molecular graphs representing the bicyclic compounds can be written as $\mathbf{A} = \mathbf{A}_0 + \mathbf{P}$, where $\mathbf{A}_0$ is the adjacency matrix for [10]annulene and $\mathbf{P}$ is a perturbation matrix, which describe the newly formed bond. Most elements of $\mathbf{P}$ are zero. Only the $(i,j)$ and $(j,i)$ elements of $\mathbf{P}$ are non-zero when a C-C bond is formed between the $i^{\text{th}}$ and $j^{\text{th}}$ carbon atoms.

***Scheme 17.*** *Generation of bicyclic compounds from [10]annulene by forming one C-C bond, where two H atoms are removed. The generated bicyclic compounds are named [m,n], where m and n indicate the number of carbon atoms composing the left and right rings, respectively. The numbering of carbon atoms in the bicyclic compounds is not the conventional one, but follows the numbering of the carbon atoms in the original [10]annulene,*

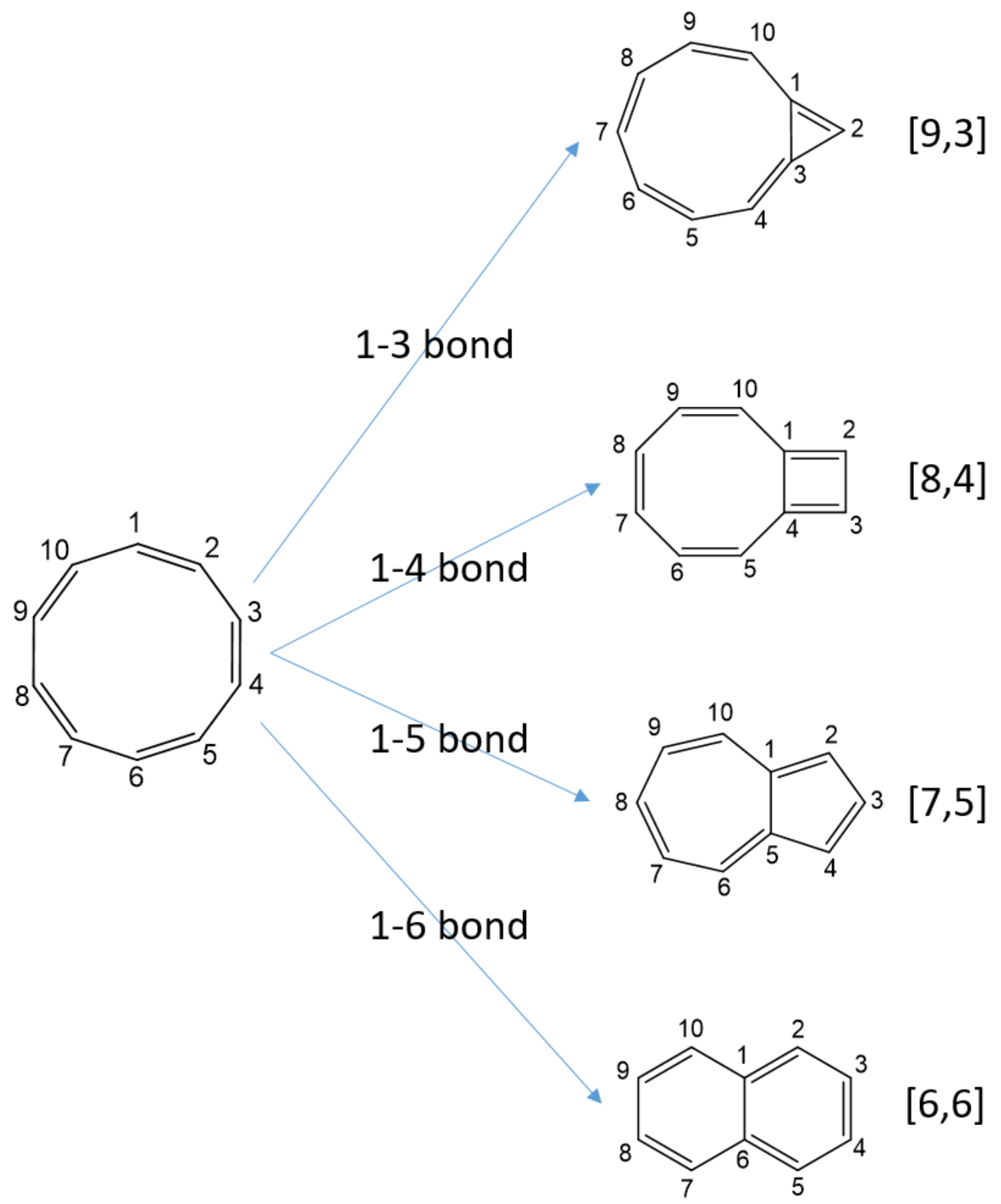

Fukui and coworkers developed a perturbation method for the secular determinant or characteristic polynomial.[137] Suppose $p_0(\lambda)$ is the characteristic polynomial of the adjacency matrix for [10]annulene. And $p_0^{i,j}(\lambda)$ is the characteristic polynomial of the adjacency matrix for [10]annulene from which the row $i$ and column $j$ are deleted. Similarly, $p_0^{ij,ij}(\lambda)$ is that of the adjacency matrix obtained by deleting the rows $i$ and $j$ and columns $i$ and $j$ from that for [10]annulene. The perturbation expression derived by Fukui and coworkers leads to the following expression:

$$p(\lambda) = p_0(\lambda) + 2(-1)^{i+j}\beta p_0^{i,j}(\lambda) - \beta^2 p_0^{ij,ij}(\lambda), \quad (62)$$

where $\beta$ is the resonance integral for the formed C$i$-C$j$ bond. We may substitute $|\beta|$ with 1.

The characteristic polynomial for the bicyclic compounds can be written in the following form: $p_{[m,n]}(\lambda) = p_0(\lambda) + p'(\lambda)$, where $p'(\lambda)$ denotes the perturbation term. Then, we obtain $p'(\lambda) = 2(-1)^{i+j}p_0^{i,j}(\lambda) - p_0^{ij,ij}(\lambda)$. Could one decide easily whether the effect of the perturbation is significant or insignificant? Let us write down an explicit form of the characteristic polynomials for [10]annulene and the various bicyclic compounds formed by bonding across the ring (see Scheme 18).

***Scheme 18.*** *Characteristic polynomials for [10]annulene ($p_0$) and bicyclic compounds related to it ($p_{[m,n]}$) are shown near the structure. The difference between $p_0$ and $p_{[m,n]}$ is regarded as the perturbation, $p'$. The red lines in the structure of [10]annulene indicate the pairs of atoms between which QI occurs. The solid blue lines in the structure of bicyclic compounds indicate the pairs of atoms between which QI occurs but did not occur between the same pair of atoms in [10]annulene. The dashed blue lines in the structure of bicyclic compounds indicate the pairs of atoms between which QI does not occur but does take place for the same pair of atoms in [10]annulene. Whether QI occurs or not is decided by checking whether a zero element appears or not in the corresponding matrix entry of the inverted adjacency matrix for each molecular graph.*

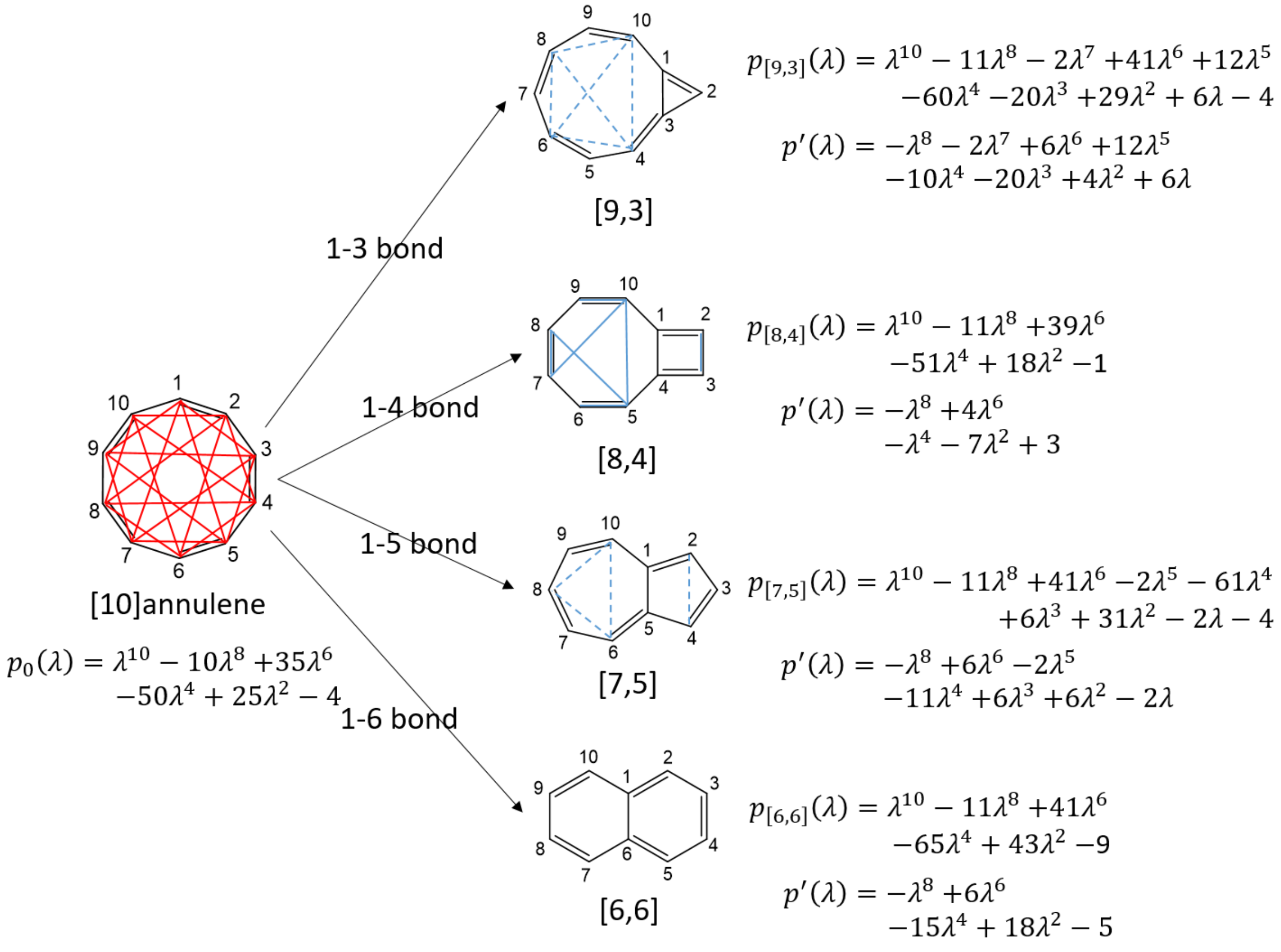

Since $p_0$ does not include any odd-power terms, all the odd-power terms in $p_{[m,n]}$ can be thought to be due to perturbation. In Fukui's formula (eq. 62), if $\beta$, the resonance integral for the formed C-C bond, is small, the perturbative term should also be small. But in our study $\beta$ is set to 1. And this cannot be said to be small. Thus, one cannot neglect the perturbation.

It is also interesting to see how QI is affected by perturbation. To this end, we compare the QI feature in [10]annulene with that in bicyclic compounds, as shown in Scheme 18. A red line connecting atoms in [10]annulene means that QI occurs between them. Most of the QI features in [10]annulene remain unchanged even if one moves from [10]annulene to a bicyclic derivative. For example, QI occurs between the $2^{nd}$ and $8^{th}$ atoms in [10]annulene, and the same is true for the connection between the $2^{nd}$ and $8^{th}$ atoms in the bicyclic compounds, [9,3], [8,4], [7,5], and [6,6]. In such a case, we do not show anything in the structure of the bicyclic compounds.

Additional QI features emerge in some of the bicyclic molecules, for instance in the [8,4] system. Such QI features are indicated by the solid blue lines. For example, there is no QI between the atoms 5 and 10 in [10]annulene, but there is in the [8,4] bicyclic molecule. Such emergence of additional QI features is actually observed only in the [8,4] bicyclic molecule.

QI features can also disappear on cross-ring bond formation, for instance in the [9,3] and [7,5] bicyclic molecules. Such disappearance of QI is indicated by the dashed blue lines. For example, there is QI between atoms 6 and 10 in [10]annulene, but there is no QI between atoms 6 and 10 in the [9,3] and [7,5] bicyclic molecules.

There is neither emergence of additional QIs nor disappearance in the [6,6] bicyclic molecule, naphthalene. Thus, one can say that the interaction between atoms 1 and 6 does not affect the QI feature, though the perturbation term $p'$ certainly does not look insignificant. We need an expression for the Green's function of an adjacency matrix which includes a perturbation (a weighted adjacency matrix).

## 11. INFINITE POWER SERIES EXPANSION OF THE GREEN'S FUNCTION

### 11.1. Neumann Series Expansion in Terms of A/E

Up to this point, we have investigated the finite series expansion of the Green's function. It may also be possible to expand the Green's function in an infinite series of the Hückel Hamiltonian or adjacency matrix. Consider the following <u>infinite</u> series:

$$\mathbf{S} = \mathbf{I} + \frac{\mathbf{A}}{E} + \frac{\mathbf{A}^2}{E^2} + \frac{\mathbf{A}^3}{E^3} + \cdots. \tag{63}$$

$E$ is bounded by the largest eigenvalue of the matrix $\mathbf{A}$. Multiplying by $\frac{\mathbf{A}}{E}$ from the left, one can obtain

$$\frac{\mathbf{A}}{E}\mathbf{S} = \frac{\mathbf{A}}{E} + \frac{\mathbf{A}^2}{E^2} + \frac{\mathbf{A}^3}{E^3} + \cdots. \tag{64}$$

**S** is not just an arbitrary expansion; taking the difference between eq. 63 and eq. 64, one obtains

$$\left(\mathbf{I} - \frac{\mathbf{A}}{E}\right)\mathbf{S} = \mathbf{I}$$
$$(E\mathbf{I} - \mathbf{A})\mathbf{S} = E\mathbf{I} \tag{65}$$
$$\mathbf{S} = [E\mathbf{I} - \mathbf{A}]^{-1}E$$
$$\mathbf{S} = \mathbf{G}E.$$

By combining eq. 65 with eq. 63, one can arrive at the following infinite power series expansion of the (energy-dependent) Green's function:

$$\mathbf{G} = [E\mathbf{I} - \mathbf{A}]^{-1} = \frac{1}{E}\left[\mathbf{I} + \frac{\mathbf{A}}{E} + \frac{\mathbf{A}^2}{E^2} + \frac{\mathbf{A}^3}{E^3} + \cdots\right]. \tag{66}$$

On the basis of this equation, one can connect the diagonal element of the Green's function to the moments of the local density of states.[58] Since $\mathbf{A}^n$ is related to the number of possible walks with length $n$ on a lattice, as detailed above, this equation implies an intimate relation between the walks and Green's function.

Two problems arise here. First, this expansion includes an infinite number of walks, making it difficult to enumerate them in practice, though the higher-order walks may not be so important, due to the larger denominator $E^n$. Second, one cannot use this equation in the limit $E \rightarrow 0$, the energy at which we assume the Fermi level is located, because the Green's function diverges. One way to avoid the problem of divergence is to add an infinitesimal imaginary number to the energy. Then, one might be able to use this expression for the description of resonant tunneling conduction. But generally one should use this equation for off-resonant conduction. We need to address the problem of convergence more carefully.

If one defines a matrix $\mathbf{M}$ as $\mathbf{A}/E$, eq. 66 can then be rewritten as

$$[\mathbf{I} - \mathbf{M}]^{-1} = \mathbf{I} + \mathbf{M} + \mathbf{M}^2 + \mathbf{M}^3 + \cdots. \tag{67}$$

This series is called a Neumann series, which is convergent if $\mathbf{M}$ is a contraction (i.e., $||\mathbf{M}|| < 1$).[138] To decide whether a Hermitian matrix is a contraction or not, one may use its spectral radius,[139] which can be calculated from

$$\rho(\mathbf{M}) = \max_k(|\lambda_k|), \tag{68}$$

where $\lambda_k$ is an eigenvalue of the matrix $\mathbf{M}$. If $\rho(\mathbf{M}) < 1$, $\mathbf{M}$ is a contraction, and the expansion converges. Since $\mathbf{M} = \mathbf{A}/E$, $\lambda_k = \varepsilon_k/E$, where $\varepsilon_k$ is an eigenvalue of $\mathbf{A}$, namely the MO energy levels. For instance, for a linear polyene consisting of $N$ carbon atoms, $\varepsilon_k$ is given in the following analytical form in the Hückel approximation,[65]

$$\varepsilon_k = \alpha + 2\beta\cos\left(\frac{k\pi}{N+1}\right), \text{ where } k = 1, 2, 3, \ldots N. \tag{69}$$

In our Hückel model, $\alpha$ is set at $E = 0$ and $\beta$ is used as the unit of energy. Therefore, the energy range of the eigenspectrum is limited to $-2 \leq \varepsilon_k \leq 2$ because of $-1 \leq \cos\theta \leq 1$, resulting in $\rho(\mathbf{M}) \leq 2/|E|$. This has to be smaller than 1 if the Neumann series is to converge. The energy range where the Neumann series expansion is valid can be $E < -2|\beta|$ and $2|\beta| < E$.

However, the QI phenomenon which we are considering appears at $E$ = 0. So it may be difficult to understand the graph-theoretic aspect of QI from the point of view of the Neumann series expansion of the Green's function. While there is a formal similarity between eq. 13 and the Neumann expansion (eq. 67), they work in different energy regimes. Eq. 13 is valid at $E = 0$, while eq. 67 is valid in the energy range of $E < -2|\beta|$ and $2|\beta| < E$.

### 11.2. Infinite Power Series Expansion of the Green's Function Based on a Perturbation Matrix

#### 11.2.1. $\mathbf{A}^{-1}$ Approach

The Green's function is the negative inverse of the adjacency matrix, so we may define a Green's function which includes a perturbation expressed by a matrix **P** as follows:

$$\mathbf{G}' = -(\mathbf{A} + \mathbf{P})^{-1}. \qquad (70)$$

Generally, we have

$$\mathbf{G}' = -(\mathbf{A} + \mathbf{P})^{-1} = -(\mathbf{I} + \mathbf{A}^{-1}\mathbf{P})^{-1}\mathbf{A}^{-1} = -\mathbf{A}^{-1} + (\mathbf{A}^{-1}\mathbf{P})\mathbf{A}^{-1} - (\mathbf{A}^{-1}\mathbf{P})^2\mathbf{A}^{-1} + \cdots. \qquad (71)$$

If one replaces $-\mathbf{A}^{-1}$ with **G**, one can recognize that this series is consistent with an infinite geometric series of the Green's function based on the Dyson equation (see SI). Eq. 71 implies that the full Green's function, which includes the effect of perturbation, can be calculated from the Green's function without the perturbation, and the perturbation matrix. Also, this equation hints at a propagation feature of the Green's function after perturbation.[140]

The series expansion as written is convergent if $||\mathbf{A}^{-1}\mathbf{P}|| < 1$. This condition may often be satisfied because the perturbation is assumed to be small. As long as the convergence criterion holds and the perturbation is small, the expansion can be truncated as follows:

$$\mathbf{G}' \approx -(\mathbf{I} - \mathbf{A}^{-1}\mathbf{P})\mathbf{A}^{-1}. \qquad (72)$$

A matrix element of eq. 72 can be explicitly written as

$$[\mathbf{G}']_{rs} \approx [-\mathbf{A}^{-1}]_{rs} + [\mathbf{A}^{-1}\mathbf{P}\mathbf{A}^{-1}]_{rs}, \qquad (73)$$

where the second term of the right-hand side of this equation can be written as

$$[\mathbf{A}^{-1}\mathbf{P}\mathbf{A}^{-1}]_{rs} = \sum_t \sum_u [\mathbf{A}^{-1}]_{rt}[\mathbf{P}]_{tu}[\mathbf{A}^{-1}]_{us}. \qquad (74)$$

If the perturbation is small, most of the elements of **P** are 0, making most of the elements of $\mathbf{A}^{-1}\mathbf{P}\mathbf{A}^{-1}$ zero.

Let us think about a perturbation due to bond formation between atoms $i$ and $j$, for example. We have already seen such a situation in Scheme 17. In this case, only the ($i$,$j$) and ($j$,$i$) elements of **P** are 1 and the other elements are 0. Thus, eq. 73 can be simplified as

$$[\mathbf{G}']_{rs} \approx [-\mathbf{A}^{-1}]_{rs} + [-\mathbf{A}^{-1}]_{ri}[-\mathbf{A}^{-1}]_{js} + [-\mathbf{A}^{-1}]_{rj}[-\mathbf{A}^{-1}]_{is}\ . \qquad (75)$$

Note that $-\mathbf{A}^{-1}$ is equivalent to the Green's function for the unperturbed system. Thus, the first term of the right-hand side of eq. 75 implies electron transport from the $r^{\text{th}}$ atom to the $s^{\text{th}}$ atom in the unperturbed system. The second term can be thought of as indicating electron transport in the unperturbed system from the $r^{th}$ atom to the $i^{th}$ atom, namely one side of the perturbation, and then that to the $s^{th}$ atom from the $j^{th}$ atom, namely the other side of the perturbation. Similarly, the last term of the right-hand side of eq. 75 implies electron transport from the $r^{th}$ atom to the $j^{th}$ atom and then that from the $i^{th}$ atom to the $s^{th}$ atom. Another, useful way to think about what happens is that the first term in eq. 75 denotes walks in the unperturbed system, while the second and third terms may be thought of as newly opened walks caused by the perturbation, namely the formation of a bond.

The series presented in eq. 71 is expanded in terms of $\mathbf{A}^{-1}$, so one may call it the "$\mathbf{A}^{-1}$ approach". By contrast, one can also expand the Green's function in terms of $\mathbf{P}^{-1}$ (see the next section).

### 11.2.2. $\mathbf{P}^{-1}$ Approach

An important assumption in the $\mathbf{P}^{-1}$ approach is that $\mathbf{P}$ has to be invertible. When $\mathbf{A}$ is an $N \times N$ matrix, $\mathbf{P}$ is also an $N \times N$ matrix. If $\mathbf{P}$ describes only one small perturbation, such as the formation of a bond between atoms $i$ and $j$, as one can see in Scheme 17, only two of the matrix elements of $\mathbf{P}$, namely $(i,j)$ and $(j,i)$, are non-zero. If we regard $\mathbf{P}$ as a kind of Hückel matrix, $\mathbf{P}$ corresponds to a molecule consisting of two bonded atoms and $N$-2 isolated atoms. Thus, $N$-2 eigenvalues of $\mathbf{P}$ are zero, so $\mathbf{P}$ is not invertible. If the perturbation is substantially more extensive, $\mathbf{P}$ may be invertible. Then, we will have

$$\mathbf{P}(\mathbf{A}+\mathbf{P})^{-1} = \left[(\mathbf{A}+\mathbf{P})\mathbf{P}^{-1}\right]^{-1} = \left[\mathbf{I}+\mathbf{A}\mathbf{P}^{-1}\right]^{-1}. \qquad (76)$$

Suppose we define $\mathbf{M} \equiv \mathbf{A}\mathbf{P}^{-1}$, the foregoing equation with the assumption that the operator norm $||\mathbf{M}|| < 1$ (is a contraction), reads

$$(\mathbf{I}+\mathbf{M})^{-1} = \mathbf{I} - \mathbf{M} + \mathbf{M}^2 - \mathbf{M}^3 + \cdots. \qquad (77)$$

Note that this is another expression of the Neumann series presented in eq. 67. If $||\mathbf{M}|| << 1$, then binominal expansion on matrices gives a good approximation:

$$(\mathbf{I}+\mathbf{M})^{-1} \approx \mathbf{I} - \mathbf{M}. \qquad (78)$$

To satisfy the convergence criterion, the perturbation must be substantially large. This condition is the same as that of judging whether $\mathbf{P}$ is invertible. Using this approximation, we have

$$\mathbf{G}' = -(\mathbf{A}+\mathbf{P})^{-1} = -\mathbf{P}^{-1}\left[\mathbf{I}+\mathbf{M}\right]^{-1} \approx -\mathbf{P}^{-1}\left[\mathbf{I}-\mathbf{M}\right] = -\mathbf{P}^{-1} + \mathbf{P}^{-1}\mathbf{A}\mathbf{P}^{-1}. \qquad (79)$$

Generally, this equation can be seen as the one obtained by switching $\mathbf{A}$ and $\mathbf{P}$ in eq. 72. In such a case, the transport properties described by **G'** are likely to be governed by the perturbation term $\mathbf{P}$ rather than the unperturbed adjacency matrix $\mathbf{A}$.

Note that we do not require Hermitian symmetry to derive the formulae presented in this section and the last section. Thus, even if **A** is the adjacency matrix of a directed graph (digraph), which is a graph having at least one directed edge (arc),[141,142] the formulations shown above hold. Directed graphs are useful for describing the network of hydrogen-bonds in water clusters, where the directed edge corresponds to hydrogen bonds from proton-donor to proton-acceptor.[143] The Green's function for a directed graph deserves consideration in the context of unsymmetrical transport, which can be found in molecular rectifiers or diodes consisting of a pair of electron-donor and acceptor units.[144,145,146] Again, this is a subject worth pursuing.

One could imagine a situation where **A** is an Hermitian adjacency matrix but **P** is non-Hermitian. One can also use the formalism presented above here as well. In this case, the Hamiltonian matrix including a perturbation, namely **A** + **P**, is usually associated with complex eigenvalues, though sometimes non-Hermitian Hamiltonians lead to real eigenvalues.[147] The complex eigenvalues totally make sense in electron transport calculations based on the non-equilibrium Green's function (NEGF) formalism.[1,49] In the NEGF method, the Hamiltonian matrix for the whole system can be divided into the central molecular region and a so-called self-energy term, which describes the interaction between the molecule and the electrode surface. The self-energy term can be regarded as the perturbation term **P** in our formalism, and is non-Hermitian, leading to a complex eigenvalue. A general interpretation of the imaginary part of the complex eigenvalues is the lifetime of electrons injected into the molecule from the electrode.[1,49]

### 11.3. An Expansion Based on the Binominal Theorem

Using the binominal theorem, Estrada and Benzi[148] found that the energy of any graph, whether bipartite or not, can be expressed as a weighted sum of the traces of even powers of the adjacency matrix as follows:

$$E = \mathrm{tr}|\mathbf{A}| = \lambda_1 \mathrm{tr}\left[\mathbf{I} + \frac{1}{2}\left(\frac{\mathbf{A}^2}{\lambda_1^2} - \mathbf{I}\right) - \frac{1}{4!!}\left(\frac{\mathbf{A}^2}{\lambda_1^2} - \mathbf{I}\right)^2 + \frac{3!!}{6!!}\left(\frac{\mathbf{A}^2}{\lambda_1^2} - \mathbf{I}\right)^3 - \cdots\right], \tag{80}$$

where $|\mathbf{A}| = \sqrt{\mathbf{A}^2}$ and $\lambda_1$ is the largest eigenvalue of **A**, which is introduced so that this expansion converges. For an even integer *n*, the double factorial (symbol !!) is the product of all even integers less than or equal to *n* but greater than or equal to 2. For an odd integer *p*, the double factorial is the product of all odd integers less than or equal to *p* and greater than or equal to 1.

By squaring eq. 80, we have

$$\mathbf{A}^2 = \lambda_1^2\left[\mathbf{I} + \frac{1}{2}\left(\frac{\mathbf{A}^2}{\lambda_1^2} - \mathbf{I}\right) - \frac{1}{4!!}\left(\frac{\mathbf{A}^2}{\lambda_1^2} - \mathbf{I}\right)^2 + \frac{3!!}{6!!}\left(\frac{\mathbf{A}^2}{\lambda_1^2} - \mathbf{I}\right)^3 - \cdots\right]^2. \tag{81}$$

Then, one can expand every squared term inside the square bracket as follows:

$$\mathbf{A}^2 = \left[a_0\mathbf{I} + a_2\mathbf{A}^2 + a_4\mathbf{A}^4 + a_6\mathbf{A}^6 + \cdots\right]^2. \qquad (82)$$

Note that the right-hand side of eq. 82 includes only even powers of **A** and **I** but their coefficients seem difficult to obtain, so they are tentatively expressed by $a_0$, $a_2$, $a_4$ and so on. One can expand the whole square in eq. 82 and obtain

$$\begin{aligned}
\mathbf{A}^2 &= \left[a_0\mathbf{I} + a_2\mathbf{A}^2 + a_4\mathbf{A}^4 + a_6\mathbf{A}^6 + \cdots\right]^2 \\
&= (a_0\mathbf{I})^2 + a_2\mathbf{A}^2\left(2a_0\mathbf{I} + a_2\mathbf{A}^2\right) + a_4\mathbf{A}^4\left(2a_0\mathbf{I} + 2a_2\mathbf{A}^2 + a_4\mathbf{A}^4\right) + a_6\mathbf{A}^6\left(2a_0\mathbf{I} + 2a_2\mathbf{A}^2 + 2a_4\mathbf{A}^4 + a_6\mathbf{A}^6\right) + \cdots \\
&= a_0{}^2\mathbf{I} + 2a_2a_0\mathbf{A}^2 + a_2{}^2\mathbf{A}^4 + 2a_4a_0\mathbf{A}^4 + 2a_4a_2\mathbf{A}^6 + a_4{}^2\mathbf{A}^8 + 2a_6a_0\mathbf{A}^6 + 2a_6a_2\mathbf{A}^8 + 2a_6a_4\mathbf{A}^{10} + a_6{}^2\mathbf{A}^{12} + \cdots \\
&= a_0{}^2\mathbf{I} + 2a_2a_0\mathbf{A}^2 + \left(a_2{}^2 + 2a_4a_0\right)\mathbf{A}^4 + \left(2a_4a_2 + 2a_6a_0\right)\mathbf{A}^6 + \cdots \\
&= \sum_{k=0}^{\infty}\sum_{j=0}^{k}\left(a_{2j}a_{2k-2j}\right)\mathbf{A}^{2k}. \quad (83)
\end{aligned}$$

One can rewrite eq. 83 as

$$\begin{aligned}
-\mathbf{I} &= -\frac{1}{a_0^2}\mathbf{A}^2 + \sum_{k=1}^{\infty}\sum_{j=0}^{k}\frac{a_{2j}a_{2k-2j}}{a_0^2}\mathbf{A}^{2k} \\
\mathbf{G} &= -\frac{1}{a_0^2}\mathbf{A} + \sum_{k=1}^{\infty}\sum_{j=0}^{k}\frac{a_{2j}a_{2k-2j}}{a_0^2}\mathbf{A}^{2k-1}.
\end{aligned} \qquad (84)$$

In the last line $\mathbf{A}^{-1}$ is multiplied from the left to obtain the Green's function.

One can see a formal similarity between this equation and eq. 13, but they are derived in a different way. The most significant difference is that eq. 13 is a finite series and only applicable to alternant hydrocarbons, while eq. 84 is an infinite series applicable to both alternant and non-alternant hydrocarbons. This equation may seem to contradict equations shown in Scheme 6 and their general expression (eq. 15), where both odd- and even-power terms are included. But this difference can be attributed to the difference between the finite and infinite character of the expressions.

Eq. 84 would imply that if there are only even-length walks between a pair of atoms, QI is expected to occur whether the molecule is alternant or non-alternant. Unfortunately, however, the simple selection rule does not work in non-alternant hydrocarbons. In non-alternant hydrocarbons, one can see a frustration of the starring scheme, where two starred (unstarred) atoms are adjacent (see Scheme 1a). If one moves from a starred (unstarred) atom to another starred (unstarred) atom without passing through the frustration region, one always takes a walk with an even-numbered length. However, once one passes through the frustration region, this does not hold true due to the adjacent existence of starred (unstarred) atoms, leading to a walk with an odd-numbered length. Thus, there are always both even- and odd-length walks between two atoms in non-alternant hydrocarbons.

## 12. SUMMARY AND CONCLUSIONS

In this work, we have made an effort to trace the origin of quantum interference, i.e.,

significant suppression of molecular conductance, back to molecular graphs and walks on them. Using the electron-hole symmetry in Hückel energy spectra coming from the pairing theorem in alternant hydrocarbons, as well as the Cayley-Hamilton theorem, we have derived an expression relating the inverse of the vertex adjacency matrix to a finite power series of the vertex adjacency matrix, where only odd-power terms are included. This result is of primary importance in this work.

Since the inverse of the adjacency matrix has an intimate relation with the Green's function, which dictates the conductive properties of molecules, the zero entries of the inverse matrix indicate the atom pair(s) between which quantum interference occurs. Since the power of the adjacency matrix is related with the number of walks on the graph, we arrive at the conclusion that only odd-length walks play an important role in molecular conductance. If there are no odd-length walks between a pair of atoms, the conductance between them is expected to be small, as a result of quantum interference. In addition, we have clarified the situation where a cancellation between the contributions from some odd-length walks leads to a zero matrix element in the inverse of the adjacency matrix. This situation is what we call "hard zero" quantum interference.

The same approach has been applied to nonalternant hydrocarbons, resulting in a power series expansion of the Green's function, where both odd- and even-power terms are included. Quantum interference in nonalternants occurs in some circumstances as a result of a cancellation between the contributions from even- and odd-length walks. This situation is reminiscent of the "hard zero" quantum interference for alternant hydrocarbons.

The finite power series expansion of the Green's function that we derive is a well-established result with straightforward applicability, immediately ready for use by the reader in his/her research. The origin of the coefficients of the power series remains a difficult problem – we feel we are just beginning to get physical and chemical insight into this aspect. This is very much work in progress and for the future.

In addition, we have explored the Green's function expansion in the form of infinite series on the basis of the Neumann series. In the case of infinite series, we often face a problem of convergence, so it may be helpful to introduce some perturbation terms to avoid singularities. We have succeeded in obtaining an infinite series by using the binominal theorem, which includes only odd-power terms and is applicable to both alternant and non-alternant hydrocarbons. However, in non-alternant hydrocarbons, if one choses two atomic sites, there are always both even-numbered-length walks and odd-numbered-length walks between them, complicating the analysis of QI in them. We feel that we are on the way to a perturbation theory of QI, in an approach that is thoroughly integrated with the powerful results of graph theory.

## ASSOCIATED CONTENT

**Supporting Information**

Understanding the Green's function based on Feynman paths, a proof for the property that the ($r$,$s$) element of $\mathbf{A}^k$ counts the number of walks of length $k$ between the nodes $r$ and $s$, explicit matrix elements of powers of the adjacency matrix for azulene, and bond and pairwise bond contributions to the square of the orbital energies for hexatriene.

**AUTHOR INFORMATION**

**Corresponding Author**

*E-mail: rh34@cornell.edu.

**ORCID**

Yuta Tsuji: 0000-0003-4224-4532

Ernesto Estrada: 0000-0002-3066-7418

Roald Hoffmann: 0000-0001-5369-6046

**Notes**

The authors declare no competing financial interests.

**Biographies**

**Yuta Tsuji** received his Ph.D. degree from Kyushu University in 2013 under the supervision of Prof. Kazunari Yoshizawa. He then worked with Prof. Roald Hoffmann at Cornell University as a JSPS fellow from 2013 to 2016. He became a research assistant professor at the Institute for Materials Chemistry and Engineering (IMCE), Kyushu University, in April 2016. He moved to the Education Center for Global Leaders in Molecular Systems for Devices in the same university as an assistant professor in August 2016. In January 2018, he returned to IMCE as an assistant professor. His research interests include molecular electronics, surface chemistry and solids.

**Ernesto Estrada** received his Ph.D. degree from the Central University of Las Villas in Cuba under the supervision of Prof. Luis A. Montero. After post-doctoral visits to the University of Valencia, Spain and the Hebrew University of Jerusalem, Israel, he moved to the University of Santiago de Compostela in Spain where he held several positions. Since 2008 he holds the Chair in Complexity Science and the 1964 Chair of Mathematics at the University of Strathclyde in Glasgow, U.K. His passion is the study of discrete complex systems ranging from molecules to large ecosystems. A meeting with Prof. Hoffmann in Glasgow some years ago triggered his interest into the study of transmission of current through molecules and the role of walks on this and other molecular properties.

**Ramis Movassagh** is a Research Staff Member at IBM Research working on the theory of quantum computing. He first joined IBM Research as a Herman Goldstine fellow in the department of mathematics. He finished his Ph.D. in mathematics at MIT. Before that he spent two and a half years at ETH-Zurich conducting research on mathematical physics and neuroscience. He received his B.Sc. in applied and engineering physics from Cornell University.

**Roald Hoffmann** received his Ph. D, with Martin P. Gouterman and William N. Lipscomb in 1962, and has been at Cornell University for 52 years. He has thought about and taught, through his papers, the ways that electrons in molecules and extended structures direct the structures, properties and reactivity of molecules. He has been fortunate to have postdoctoral fellows who have drawn him into fields of chemistry and physics that he never imagined he would think about. It was Yuta Tsuji who in this way enticed him in to consider transmission of current through molecules.

## ACKNOWLEDGMENTS

We are grateful to S. Datta and M. H. Garner for their comments on this work, and the reviewers for their suggestions, especially of a summary paragraph for section 5, and the inclusion of a Reader's Guide. Y.T. thanks the Research Institute for Information Technology (Kyushu University) for the computer facilities and financial support from JSPS KAKENHI Grant Number JP17K14440 and from Qdai-jump Research Program, Wakaba Challenge of Kyushu University. The work at Cornell was supported by the National Science Foundation through Grant CHE 1305872. E.E. thanks the Royal Society of London for a Wolfson Research Merit Award. R.M. thanks IBM TJ Watson Research for the freedom and the support.

**TOC graphic**

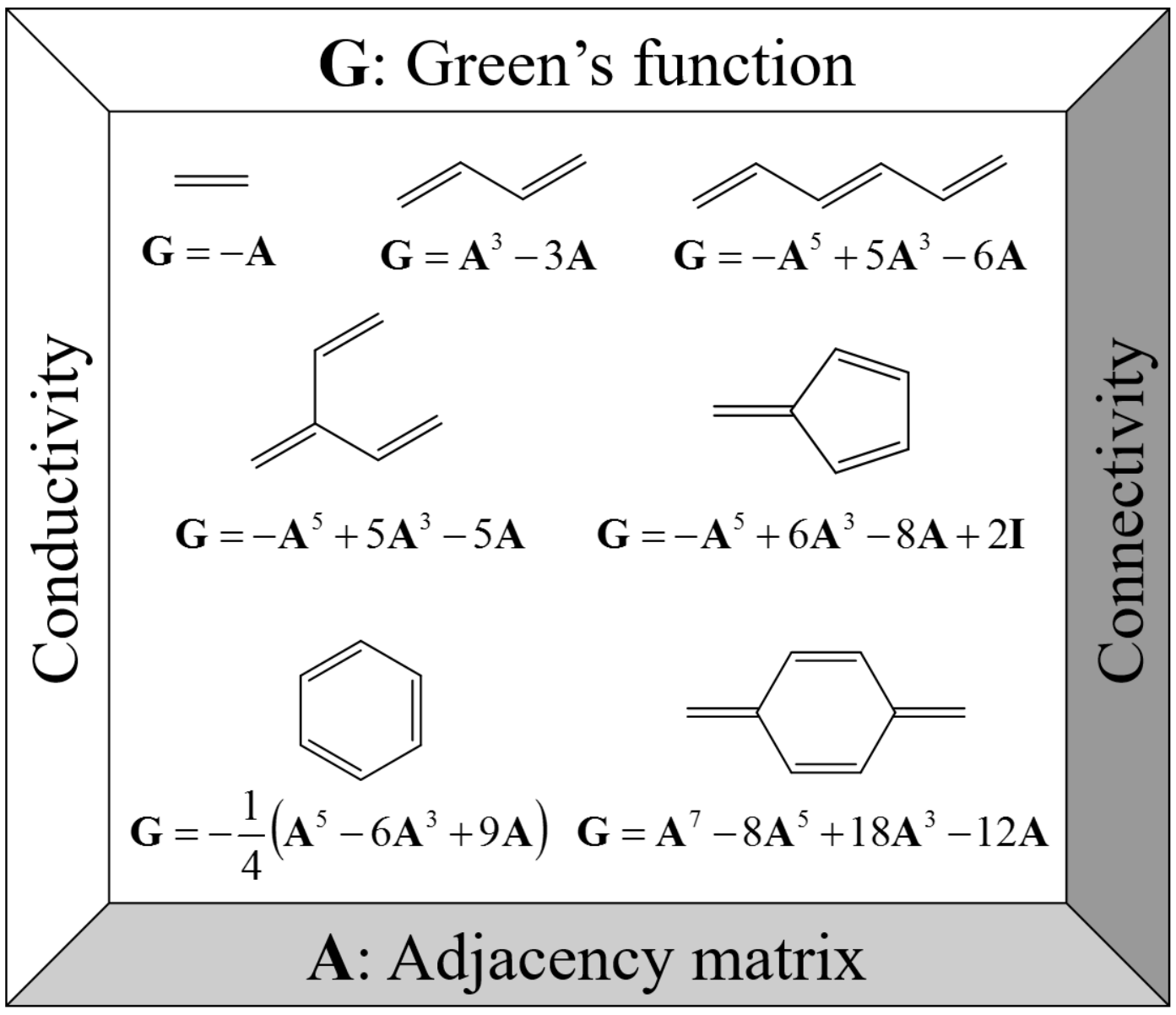